\documentclass[nofootinbib,showpacs,preprint,preprintnumbers,amsmath,amssymb]{revtex4-1}
\pdfoutput=1
\usepackage[caption=false]{subfig}
\usepackage{bm}
\usepackage{amsmath}
\usepackage{amsfonts}
\usepackage{amssymb}
\usepackage{float}
\usepackage{color}
\usepackage{graphicx}
\usepackage{graphics}
\usepackage{hyperref}
\usepackage{adjustbox,array}
\usepackage{enumitem} 
\hypersetup{}
\usepackage[utf8x]{inputenc} 
\usepackage{epstopdf}
\usepackage{multirow}
\usepackage{slashed}
\usepackage{booktabs}
\usepackage{cancel}
\usepackage{mathrsfs}
\usepackage{csquotes}

\DeclareUnicodeCharacter{2217}{}
\DeclareUnicodeCharacter{2212}{}

\definecolor{rosso}{cmyk}{0,1,1,0.4}
\definecolor{rossos}{cmyk}{0,1,1,0.55}
\definecolor{rossoc}{cmyk}{0,1,1,0.2}
\definecolor{blu}{cmyk}{1,1,0,0.3}
\definecolor{blus}{cmyk}{1,1,0,0.6}
\definecolor{bluc}{cmyk}{1,1,0,0.1}
\definecolor{verde}{cmyk}{0.92,0,0.59,0.25}
\definecolor{verdec}{cmyk}{0.92,0,0.59,0.15}
\definecolor{verdes}{cmyk}{0.92,0,0.59,0.4}

\AtBeginDocument{
\heavyrulewidth=.08em
\lightrulewidth=.05em
\cmidrulewidth=.03em
\belowrulesep=.65ex
\belowbottomsep=0pt
\aboverulesep=.4ex
\abovetopsep=0pt
\cmidrulesep=\doublerulesep
\cmidrulekern=.5em
\defaultaddspace=.5em
}

\hypersetup{
 colorlinks=true,
 linkcolor=verdec,
 urlcolor=verdec,
 citecolor=verdec
}

\begin{document}

\title{\color{verdes} Probing Zee-Babu states at Muon Colliders}

\author{Adil Jueid}
\email{adiljueid@ibs.re.kr}
\address{Particle Theory and Cosmology Group, Center for Theoretical Physics of the Universe,
Institute for Basic Science (IBS), 34126 Daejeon, Republic of Korea}

\author{Talal Ahmed Chowdhury}
\email{talal@du.ac.bd}
\address{Department of Physics, University of Dhaka, Dhaka 1000, Bangladesh}
\address{Department of Physics and Astronomy, University of Kansas, Lawrence, Kansas 66045, USA}
\address{The Abdus Salam International Centre for Theoretical Physics, Strada Costiera 11, I-34014, Trieste, Italy}
\author{Salah Nasri}
\email{snasri@uaeu.ac.ae}
\email{salah.nasri@cern.ch}
\address{Department of physics, United Arab Emirates University, Al-Ain, UAE}
\address{The Abdus Salam International Centre for Theoretical Physics, Strada Costiera 11, I-34014, Trieste, Italy}

\author{Shaikh Saad}
\email{shaikh.saad@unibas.ch}
\address{Department of Physics, University of Basel, Klingelbergstrasse\ 82, CH-4056 Basel, Switzerland}

\begin{abstract}
The Zee-Babu model is a minimal realization of radiative neutrino mass generation mechanism at the two-loop level. We study the phenomenology of this model at future multi-TeV muon colliders. After imposing all theoretical and low-energy experimental constraints on the model parameters, we find that the Zee-Babu states are expected not to reside below the TeV scale, making it challenging to probe them at the LHC. We first analyze the production rates for various channels, including multi singly-charged and/or doubly-charged scalars at muon colliders. For concreteness, we study several benchmark points that satisfy neutrino oscillation data and other constraints and find that most channels have large production rates. We then analyze the discovery reach of the model using two specific channels: the pair production of singly- and doubly-charged scalars. For the phenomenologically viable scenarios considered in this study, charged scalars with masses up to ${\cal O}(3$--$4)$ TeV can be probed for the center-of-mass energy of $10$ TeV and total luminosity of $10~{\rm ab}^{-1}$. 
\end{abstract}

\vspace{1cm}
\keywords{Neutrino Mass, Muon Colliders}

\preprint{CTPU-PTC-23-23}

\maketitle
\tableofcontents

\section{Introduction}

Although the Standard Model (SM) of particle physics is the most successful theory to describe nature at the fundamental scale, it has several drawbacks. Among its shortcoming, the most prominent is that neutrinos are massless in the SM. However, several experiments~\cite{Super-Kamiokande:1998kpq,Super-Kamiokande:2001ljr,SNO:2002tuh,KamLAND:2002uet,KamLAND:2004mhv,K2K:2002icj,MINOS:2006foh} have discovered neutrino oscillations that firmly established non-zero neutrino masses for at least two generations. Observations of non-zero neutrino masses unquestionably call for physics beyond the SM (BSM). A well-motivated class of models explaining the origin and the smallness of the neutrino mass is the radiative neutrino mass~\cite{Cheng:1977ir} generation mechanism where the neutrino mass is generated at one or more loops with the BSM particles whose masses are typically not too above the TeV scale. An economical way of generating neutrino masses, which only extends the scalar sector of the SM, are the Zee model~\cite{Zee:1980ai,Zee:1985id} and the Zee-Babu model~\cite{Cheng:1980qt,Babu:1988ki}. In the former case, a singly-charged scalar and  a second Higgs doublet are introduced, and neutrino masses appear at the one-loop level. The latter model extends the SM scalar sector by only a singly-charged scalar  and a doubly-charged scalar, and, consequently,    neutrino masses arise at two-loop order. This model has been widely studied in the literature~\cite{Babu:2002uu,AristizabalSierra:2006gb,Nebot:2007bc,Ohlsson:2009vk,Schmidt:2014zoa,Herrero-Garcia:2014hfa,Babu:2015ajp} (see also Ref. \cite{Cai:2017mow} for a comprehensive review), and different versions are also proposed. For example, colored versions of the Zee-Babu model are studied in Refs.~\cite{Babu:2001ex,Babu:2010vp,Kohda:2012sr,Guo:2017gxp,Datta:2019tuj,Babu:2019mfe}.

In this work, we focus on the minimal version of the Zee-Babu model and consider the possibilities of observing the new physics states, namely, the single-charged and doubly-charged scalars at the muon colliders. 
A salient feature of this  model is that owing to two-loop suppression, obtaining the correct neutrino mass scale requires that the new physics states are not too heavy or the Yukawa couplings are not too small. For the BSM scalars of masses $\mathcal{O}(100)$ GeV, there is a lower bound of $> 10^{-2}$ on  the largest of the Yukawa couplings. Consequently, several lepton flavor violating (LFV) processes mediated by the Zee-Babu scalars can become observable. Among them, the most promising LFV process is $\mu\to e\gamma$, which cannot be arbitrarily small. As will be explained later, if fine-tuned regions of the parameter space are not considered, then non-observation of LFV signals and reproducing the observed neutrino oscillation data prefer the new physics states to reside at or above $\sim$ TeV. Moreover, a recent analysis of the LHC puts a lower limit of about TeV on the mass of the doubly-charged scalars.

Recently, multi-TeV muon colliders have attracted interest in discovering new physics at the TeV scale and beyond (for a review see Refs. \cite{Delahaye:2019omf,Long:2020wfp,AlAli:2021let,Accettura:2023ked}). One of the important aspects of multi-TeV muon colliders is that we can achieve both a high center-of-mass energy and a clean environment since the backgrounds are mostly of electroweak origin. For example, a signal-to-background ratio for Higgs boson production at muon colliders is about $10^{-2}$ for all the center-of-mass energies while it is about $10^{-6}$ at the LHC. This would imply that the multi-TeV muon colliders are the perfect colliders for discovery, precision measurements, and even new physics characterisation. Phenomenological analyses have been extensively carried out at  future muon colliders both for the SM and new physics \cite{Capdevilla:2020qel,Chiesa:2020awd,Han:2020uid,Han:2020uak,Yin:2020afe,Huang:2021nkl,Capdevilla:2021rwo,Capdevilla:2021fmj,Asadi:2021gah,Casarsa:2021rud,Liu:2021akf,Han:2021udl,Han:2021kes,Han:2021lnp,Lv:2022pts,Liu:2022byu,Azatov:2022itm,Yang:2022fhw,Bao:2022onq,Chen:2022msz,Costantini:2020stv,Ruiz:2021tdt,Homiller:2022iax,Jueid:2023zxx,Garosi:2023bvq}. In this work, we show that multi-TeV muon colliders have great potential to discover charged scalar states. One of the main reasons for this is that new physics states typically couple with muons stronger than the other generations within this framework. Moreover, the doubly-charged scalar may leave some clean signatures through which this model can be efficiently probed at future muon colliders. In particular, using a fully-fledged analysis at the detector level, we show that singly- and doubly-charged scalars of masses about $1.25$--$3.75$ TeV  can be efficiently probed at muon colliders.

The study performed in this work for searching Zee-Babu states in muon colliders may also apply to various models. Because, apart from the Zee and Zee-Babu models,  there is a plethora of radiative neutrino mass models for which charged scalars are essential ingredients.  The scotogenic model~\cite{Ma:2006km} that links dark matter to the radiative generation of neutrino masses comes with a singly-charged scalar originating from an inert doublet. Moreover, going beyond two-loop, there are three minimal three-loop neutrino mass models, namely (i) KNT model~\cite{Krauss:2002px}, (ii) AKS model~\cite{Aoki:2008av}, and (iii)
cocktail model~\cite{Gustafsson:2012vj}. The KNT model consists of two singly-charged scalars and three copies of right-handed neutrinos. The AKS model extends the SM scalar sector with a second Higgs doublet, a SM singlet, and a singly-charged scalar. In addition to the Zee-Babu states, the cocktail model introduces an inert doublet. In this model, however, the singly-charged scalar carries a discrete charge, and the Zee-Babu loop is not allowed. Another model worth mentioning is the BNT model~\cite{Babu:2009aq}, which, in addition to a singly-charged and a doubly-charged scalars also predict the existence of a triply-charged scalar. All the models mentioned so far assume neutrinos are Majorana particles; however,  neutrinos can also be Dirac. The minimal radiative Dirac neutrino mass models also predict the existence of charged scalars~\cite{Kanemura:2011jj,Ma:2014qra,Calle:2018ovc,Bonilla:2018ynb}). The above list does not provide a complete set of references, and we refer the reader to the recent review Ref.~\cite{Cai:2017jrq}.

This work is organized in the following way. After introducing the model in Sec.~\ref{sec:model}, we discuss various theoretical and experimental constraints on model parameters in Sec.~\ref{sec:constraints}. By imposing the derived constraints, we perform numerical fit to the neutrino oscillation data in Sec.~\ref{sec:numerical} and present five benchmark case studies. A detailed analysis of the production rates of new scalar states at  muon colliders is then carried out in Sec.~\ref{sec:collider} and discovery prospects for two production channels are illustrated in  Sec.~\ref{ref:sensitivity}. Finally, we conclude in Sec.~\ref{sec:conclusions}.

\section{Model}
\label{sec:model}

The details of Zee-Babu model~\cite{Cheng:1980qt,Babu:1988ki} are presented in this section. The particle content of the SM is extended with only a singly-charged and a doubly-charged scalars,
\begin{align}
&\phi^+\sim (1,1,1),\;\;\; \kappa^{++}\sim (1,1,2).   
\end{align}
The kinetic part of the Lagrangian for these two fields is given by,
\begin{align}
&\mathcal{L}^\mathrm{NP}_K=\left(D_\mu \phi\right)^\dagger   \left(D_\mu \phi\right)
+\left(D_\mu \kappa\right)^\dagger   \left(D_\mu \kappa\right),
\\
&D_\mu=\partial_\mu+i g_Y Y_x B_\mu,\;\;\;
Y_\phi=+1,\; Y_\kappa=+2.
\end{align}
The scalar potential of this theory also takes a simple form,
\begin{align}
V^\mathrm{NP}&= \mu^2_\phi |\phi|^2 +\lambda_\phi |\phi|^4 
+\mu^2_\kappa |\kappa|^2 +\lambda_\kappa |\kappa|^4
+\lambda_{\phi\kappa}|\phi|^2|\kappa|^2
+\lambda_{H\phi}|H|^2|\phi|^2
+\lambda_{H\kappa}|H|^2|\kappa|^2
+\lambda_{\phi\kappa}|\phi|^2|\kappa|^2
\nonumber
\\&
+ \left( \mu \phi^+\phi^+ \kappa^{--}   + {\rm h.c.} \right),
\end{align}
where $H\sim (1,2,1/2)$ denotes the SM Higgs. The last term in $V^\mathrm{NP}$ plays a crucial role in generating Majorana neutrino masses. 

Moreover, the Yukawa couplings associated to these new states take the following form: 
\begin{align}
\mathcal{L}^\mathrm{NP}_Y&= f_{ij} L^{aT}_i C L^b_j \epsilon_{ab} \phi^+ + g_{ij} \ell^T_i C \ell_j \kappa^{++} + {\rm h.c.}
\end{align}
Here, we have used the notation $L\sim (1,2,-1/2)\sim (\nu_L, \ell_L)^T$ and $\ell\sim (1,1,-1)=\ell_R$, and $i,j$ ($a,b$) are family (weak) indices.  The $3\times 3$ Yukawa coupling matrices $f$ and $g$ are anti-symmetric and symmetric, respectively, in the family space. Therefore, $f$ has three independent entries  $\left(f_{e\mu},f_{e\tau},f_{\mu\tau}\right)$ and $g$ has
six $\left( g_{ee}, g_{e\mu}, g_{e\tau}, g_{\mu\mu}, g_{\mu\tau}, g_{\tau\tau} \right)$.  For the simplicity of the analysis, we take all   parameters to be real.

\begin{figure}[th!]\centering
\includegraphics[width=0.7\textwidth]{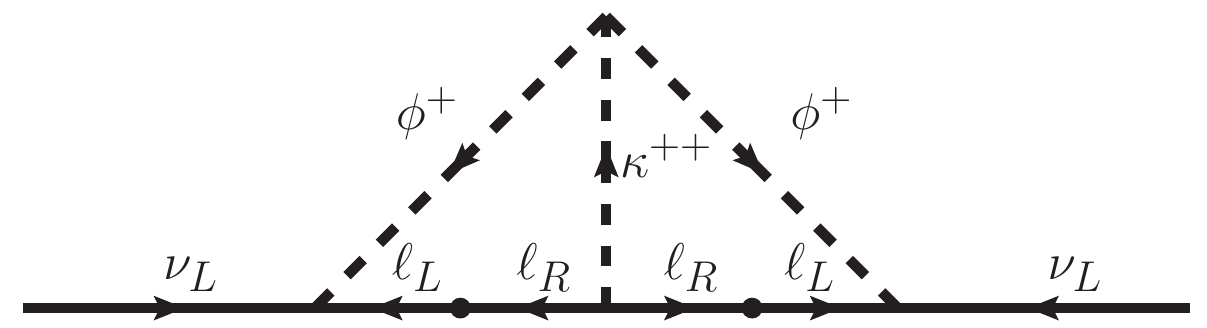}
\caption{Neutrino mass generation at the two-loop order. }\label{fig:numass}
\end{figure} 
If we assign two units of lepton number ($L$) to the BSM states, i.e., $L[\phi]=-2$, $L[\kappa]=-2$, then  all terms in the entire Lagrangian conserve lepton number  except the term $\phi^+\phi^+ \kappa^{--}$ in the scalar potential. Consequently, lepton number is broken by two units and neutrinos acquire Majorana mass at the two-loop level as depicted in figure~\ref{fig:numass}.

This two-loop diagram is calculable, and 
neutrino mass formula takes the following form~\cite{Nebot:2007bc}:  
\begin{align}
\mathcal{M}^\nu_{ij}=16\mu f_{ik}m_k  g^\ast_{kl} m_l f_{lj} I_{kl}, \label{eq:numass}
\end{align}
where charged lepton masses are denoted by $m_i$. Since the masses of the charged leptons  are much smaller than the charged scalar masses, they can be neglected from the corresponding propagators in the loop integral. Then   the corresponding loop function becomes essentially generation independent, and is given by, 
\begin{align}
&I_{kl}\approx I= \frac{1}{(16\pi^2)^2} \frac{1}{m^2_\phi} \overline I\left[ \frac{m^2_\kappa}{m^2_\phi} \right], 
\end{align}
with
\begin{align}
\overline I\left[x\right] =-\int^1_0 dz \int^{1-z}_0 dy \frac{1-y}{z+(x-1)y+y^2} \log \frac{y(1-y)}{z+xy}.    \label{eq:int} 
\end{align}
In some special cases, the loop function takes the following simple forms:
\begin{equation}
  \setlength{\arraycolsep}{0pt}
  \overline I\left[ x \right]\longrightarrow  \left\{ \begin{array}{ l l }
    \frac{\log^2x+\pi^2/3-1}{x},\;\;\; x\gg 1 \\
    \pi^2/3,\hspace{45pt} x\to 0. 
  \end{array} \right.
\end{equation}
For numerical analysis, we evaluate the integral given in equation~\eqref{eq:int}, and use the following expression for the neutrino mass matrix:
\begin{align}
\mathcal{M}^\nu= \frac{16\mu\; \overline I\left[ m^2_\kappa/m^2_\phi \right]}{(16\pi^2)^2 \;m^2_\phi}
f m^\mathrm{diag}_E  g^\dagger m^\mathrm{diag}_E f^T. \label{eq:numass:matrix}
\end{align}
In the above equation, $m^\mathrm{diag}_E$ represents the diagonal charged lepton mass matrix, namely $m^\mathrm{diag}_E=\mathrm{diag}(m_e, m_\mu, m_\tau)$.

Note that due to the antisymmetric nature of the Yukawa matrix $f$, the determinant of the neutrino mass matrix vanishes. Therefore, lightest of the neutrinos remain massless at two-loop order.  
In general, the neutrino mass matrix given in equation~\eqref{eq:numass} can satisfy both the normal and inverted mass ordering. Neutrino oscillation data for these two cases are collected in Table~\ref{tab:nuEXP}.  
\begin{table}[th!] 
\centering
\begin{footnotesize}
    \begin{tabular}{c|l|cc|cc}
      \hline\hline
      \multirow{11}{*}{} &
      & \multicolumn{2}{c|}{Normal Ordering}
      & \multicolumn{2}{c}{Inverted Ordering}
      \\
      \cline{3-6}
      && bfv $\pm 1\sigma$ & $3\sigma$ range
      & bfv $\pm 1\sigma$ & $3\sigma$ range
      \\
      \cline{2-6}
      \rule{0pt}{4mm}\ignorespaces
      & $\sin^2\theta_{12}$
      & $0.304_{-0.012}^{+0.013}$ & $0.269 \to 0.343$
      & $0.304_{-0.012}^{+0.012}$ & $0.269 \to 0.343$
      \\[3mm]
      & $\sin^2\theta_{23}$
      & $0.573_{-0.023}^{+0.018}$ & $0.405 \to 0.620$
      & $0.578_{-0.021}^{+0.017}$ & $0.410 \to 0.623$
      \\[3mm]
      & $\sin^2\theta_{13}$
      & $0.02220_{-0.00062}^{+0.00068}$ & $0.02034 \to 0.02430$
      & $0.02238_{-0.00062}^{+0.00064}$ & $0.02053 \to 0.02434$
      \\[3mm]
      & $\frac{\Delta m^2_{21}}{10^{-5}}$ eV$^2$
      & $7.42_{-0.20}^{+0.21}$ & $6.82 \to 8.04$
      & $7.42_{-0.20}^{+0.21}$ & $6.82 \to 8.04$
      \\[3mm]
      & $\frac{\Delta m^2_{3\ell}}{10^{-3}}$ eV$^2$
      & $2.515_{-0.028}^{+0.028}$ & $2.431 \to 2.599$
      & $-2.498_{-0.029}^{+0.028}$ & $-2.584 \to -2.413$
      \\[2mm]
      \hline\hline
      \end{tabular}
\end{footnotesize}
\caption{Neutrino oscillation parameters taken from Ref.~\cite{Esteban:2020cvm}. Here, $\Delta m^2_{31} > 0$ for NH and
    $\Delta m^2_{32} < 0$ for IH. Here `bfv' represents best fit values obtained from global fit~\cite{Esteban:2020cvm}. }\label{tab:nuEXP}
\end{table}

Interestingly, due to the two-loop suppression, the larger of the couplings have to be $\gtrsim 10^{-2}$ to reproduce the correct neutrino mass scale. Consequently, this model can be efficiently probed through the rare charged lepton flavor violating processes.  For analysing the implications in muon colliders, we are particularly interested in coupling of order $\sim \mathcal{O}(1)$, especially the $\mu\mu$ coupling to new physics states. Before we perform a fit to the neutrino masses and mixings, in the next section, we summarize the theoretical and experimental constraints on the model parameters.

\section{Theoretical and experimental constraints}
\label{sec:constraints}

In this section, we summarize various theoretical as well as experimental constraints on the model parameters.

\subsection{Theoretical constraints} 
Model parameters that play role in neutrino mass generation are the Yukawa couplings $f_{ij}, g_{ij}$ and the cubic coupling $\mu$. Therefore, we are interested in finding the constraints on these parameters.  First, due to the perturbativity of the Yukawa couplings, we restrict ourselves to   $f_{ij}, g_{ij}\leq 1$. As for the cubic coupling, the most crucial bound arises from loop effects. In particular, due to the cubic coupling $\phi^+\phi^+ \kappa^{--}$, effective quartic interactions for $\phi^+$ and $\kappa^{++}$ are generated at one-loop level. These effective quartic coupling are negative, and with $m_\phi, m_\kappa$ masses of similar order,  they are given by $\lambda_\mathrm{eff}\sim -\mu^4/(6\pi^2m^4_\phi)$~\cite{Babu:2002uu}.  Hence, the tree-level quartic couplings ($\lambda_{\phi}, \lambda_{\kappa}, \lambda_{\phi\kappa}$) must be positive and larger in magnitude such that the net effective quartic couplings are positive to ensure the vacuum stability. Then, assuming perturbitivity and restricting to values $\lambda_{\phi}, \lambda_{\kappa}, \lambda_{\phi\kappa}\leq 1$, constraints on the cubic coupling parameter is obtained~\cite{Babu:2002uu}, which read
\begin{equation}
  \setlength{\arraycolsep}{0pt}
  \mu\leq   \left\{ \begin{array}{l l l}
    m_\phi\times \left(6\pi^2\right)^{1/4},\;\;\; m_\kappa\approx m_\phi. \\
    m_\phi\times \left(2\pi^2\right)^{1/4},\;\;\; m_\kappa\ll m_\phi. \\
    m_\phi\times \left(24\pi^2\right)^{1/4},\;\;\; m_\kappa\gg m_\phi. 
  \end{array} \right. \label{eq:mu-1}
\end{equation}
Since all the cases we study are such that $m_\kappa$ and $m_\phi$ are closer in mass, while performing a numerical analysis, we impose the first constraint listed above.

As will be shown in the next section, a fit to the neutrino oscillation data prefers somewhat larger values of $\mu$. Such a large value compared to the masses of the scalars can be dangerous if it leads to a deeper minimum of the scalar potential for non-vanishing values of charged scalars leading to charge-breaking minimum. One must impose relevant constraints to avoid charge breaking minimum, which, for example, are studied in Refs.~\cite{Frere:1983ag,Alvarez-Gaume:1983drc,Casas:1996de}. By applying similar method to Zee-Babu model, a conservative bound on the cubic coupling is obtained in Ref.~\cite{Herrero-Garcia:2014hfa},
\begin{align}
\mu \lesssim 8\; \mathrm{max}\left(m_\phi, m_\kappa\right).    \label{eq:mu-2}  
\end{align}
However, a much more restrictive bound on this parameter is suggested in Ref.~\cite{Herrero-Garcia:2014hfa} (see also Ref.~\cite{Schmidt:2014zoa}), which is given by,
\begin{align}
\mu \lesssim 4\pi\; \mathrm{min}\left(m_\phi, m_\kappa\right).    \label{eq:mu-3} 
\end{align}
In our numerical analysis, we make sure that all these conditions, equation~\eqref{eq:mu-1} (the first line), equation~\eqref{eq:mu-2}, and equation~\eqref{eq:mu-3} are met.

\begin{table}[t!]
\centering
\begin{tabular}{ccl}
\hline 
Process&
Experiment (90\% CL)&
Bound (90\% CL)\tabularnewline
\hline
$\mu^{-}\rightarrow e^{+}e^{-}e^{-}$&
BR$<1.0\times10^{-12}$&
$| g_{e\mu} g_{ee}^{*}|/m^2_\kappa< 2.33\times 10^{-11}$ GeV$^{-2}$ \tabularnewline
$\tau^{-}\rightarrow e^{+}e^{-}e^{-}$&
BR$<2.7\times10^{-8}$&
$| g_{e\tau} g_{ee}^{*}|/m^2_\kappa< 9.07\times 10^{-9}$ GeV$^{-2}$\tabularnewline
$\tau^{-}\rightarrow e^{+}e^{-}\mu^{-}$&
BR$<1.8\times10^{-8}$&
$| g_{e\tau} g_{e\mu}^{*}|/m^2_\kappa< 5.23\times 10^{-9}$ GeV$^{-2}$\tabularnewline
$\tau^{-}\rightarrow e^{+}\mu^{-}\mu^{-}$&
BR$<1.7\times10^{-8}$&
$| g_{e\tau} g_{\mu\mu}^{*}|/m^2_\kappa< 7.20\times 10^{-9}$ GeV$^{-2}$\tabularnewline
$\tau^{-}\rightarrow\mu^{+}e^{-}e^{-}$&
BR$<1.5\times10^{-8}$&
$| g_{\mu\tau} g_{ee}^{*}|/m^2_\kappa< 6.85\times 10^{-9}$ GeV$^{-2}$\tabularnewline
$\tau^{-}\rightarrow\mu^{+}e^{-}\mu^{-}$&
BR$<2.7\times10^{-8}$&
$| g_{\mu\tau} g_{e\mu}^{*}|/m^2_\kappa< 6.50\times 10^{-9}$ GeV$^{-2}$\tabularnewline
$\tau^{-}\rightarrow\mu^{+}\mu^{-}\mu^{-}$&
BR$<2.1\times10^{-8}$&
$| g_{\mu\tau} g_{\mu\mu}^{*}|/m^2_\kappa< 8.11\times 10^{-9}$ GeV$^{-2}$\tabularnewline
$\mu^{+}e^{-}\rightarrow\mu^{-}e^{+}$&
$G_{M\bar{M}}<0.003\times G_{F}$&
$| g_{ee} g_{\mu\mu}^{*}|/m^2_\kappa< 1.97\times 10^{-7}$ GeV$^{-2}$\tabularnewline
\hline
\end{tabular}
\caption{Constraints from tree-level lepton flavour violating decays. Experimental limits are taken from~\cite{BELLGARDT19881,Hayasaka:2010np}.  }\label{tab:tree}
\end{table}

\subsection{Low-energy experimental constraints}

The Yukawa couplings that give rise to neutrino masses also  participate in rare charged lepton violating (cLFV) processes.  Such cLFV processes $\ell^-_i\to \ell^+_j\ell^-_k\ell^-_l$ already take place at tree-level and $\ell^-_i\to \ell^+_j\gamma$  appears at one-loop level.  The partial widths for these decays are given by~\cite{Nebot:2007bc},
\begin{align}
&R_\Gamma(\ell^-_i\to \ell^-_j\gamma)\equiv  \frac{\Gamma(\ell^-_i\to \ell^-_j\gamma)}{\Gamma(\ell^-_i\to \ell^-_j\nu \overline \nu)}= \frac{\alpha}{48\pi G^2_F} \bigg\{ \left(\frac{(f^\dagger f)_{ij}}{m^2_\phi}\right)^2+ 16 \left(\frac{( g^\dagger  g)_{ij}}{m^2_\kappa}\right)^2 \bigg\},    
\end{align}
and,
\begin{align}
&R_\Gamma(\ell^-_i\to \ell^+_j\ell^-_k\ell^-_l)\equiv  \frac{\Gamma(\ell^-_i\to \ell^+_j\ell^-_k\ell^-_l)}{\Gamma(\ell^-_i\to \ell^-_j\nu \overline \nu)}= \frac{1}{2(1+\delta_{kl})G^2_F m^4_\kappa} |  g_{ij} g^\ast_{kl} |^2. \end{align}
In the last formula, $\delta_{kl}$ takes into account the possibility of having two identical particles in the final state. Utilizing the above formulas, the relevant branching fractions are obtained by computing ${\rm BR}(\ell^-_i\to \ell^+_j\ell^-_k\ell^-_l)=R_\Gamma(\ell^-_i\to \ell^+_j\ell^-_k\ell^-_l)\times {\rm BR}(\ell^-_i\to \ell^-_j\nu\overline \nu)$ and ${\rm BR}(\ell^-_i\to \ell^-_j\gamma)=R_\Gamma(\ell^-_i\to \ell^-_j\gamma)\times  {\rm BR}(\ell^-_i\to \ell^-_j\nu\overline \nu)$.

At one-loop, new physics contribution to the anomalous magnetic moment of the $i$-th generation of lepton reads~\cite{Nebot:2007bc}, 
\begin{align}
\delta(g-2)_i= -\frac{m^2_i}{24\pi^2}\bigg\{ \frac{\left(f^\dagger f\right)_{ii}}{m^2_\phi} +4\frac{\left( g^\dagger  g\right)_{ii}}{m^2_\kappa} \bigg\}.    
\end{align}
Furthermore, another LFV process, namely, the muonium-antimuonium transition is also mediated via the tree-level interactions.  Consequently, the following effective Lagrangian coupling coefficient is generated~\cite{Horikawa:1995ae,Nebot:2007bc}
 \begin{align}
G_{M\bar{M}}= -\frac{1}{4\sqrt{2}m^2_\kappa}  g_{11} g^\ast_{22}.   
\end{align}

\begin{table}[t!]
\centering
\begin{tabular}{cl}
\hline 
Experiment&
~~~~~Bound (90\%CL)\tabularnewline
\hline
$\delta a_{e}=2.8\times10^{-13}$&
$\frac{|f_{e\mu}|^{2}+|f_{e\tau}|^{2}}{m^2_\phi}+4\frac{|g_{ee}|^{2}+|g_{e\mu}|^{2}+| g_{e\tau}|^{2}}{m^2_\kappa}<2.53\times 10^{-4}$ GeV$^{-2}$\tabularnewline
$\delta a_{\mu}=2.61\times10^{-9}$&
$\frac{|f_{e\mu}|^{2}+|f_{\mu\tau}|^{2}}{m^2_\phi} +4\frac{| g_{e\mu}|^{2}+| g_{\mu\mu}|^{2}+| g_{\mu\tau}|^{2}}{m^2_\kappa} <5.53\times 10^{-5}$ GeV$^{-2}$\tabularnewline

${\rm BR}(\mu\rightarrow e\gamma)<4.2\times10^{-13}$&
$\frac{|f_{e\tau}^{*}f_{\mu\tau}|^{2}}{m^4_\phi}+16\frac{| g_{ee}^{*}g_{e\mu}+ g_{e\mu}^{*} g_{\mu\mu}+ g_{e\tau}^{*} g_{\mu\tau}|^{2}}{m^4_\kappa} <1.10\times 10^{-18}$ GeV$^{-4}$ \tabularnewline
${\rm BR}(\tau\rightarrow e\gamma)<3.3\times10^{-8}$&
$\frac{|f_{e\mu}^{*}f_{\mu\tau}|^{2}}{m^4_\phi}+16\frac{| g_{ee}^{*} g_{e\tau}+ g_{e\mu}^{*}g_{\mu\tau}+ g_{e\tau}^{*} g_{\tau\tau}|^{2}}{m^4_\kappa} <4.85\times 10^{-13}$ GeV$^{-4}$ \tabularnewline
${\rm BR}(\tau\rightarrow\mu\gamma)<4.4\times10^{-8}$&
$\frac{|f_{e\mu}^{*}f_{e\tau}|^{2}}{m^4_\phi}+16\frac{| g_{e\mu}^{*} g_{e\tau}+ g_{\mu\mu}^{*} g_{\mu\tau}+ g_{\mu\tau}^{*} g_{\tau\tau}|^{2}}{m^4_\kappa} <6.65\times 10^{-13}$ GeV$^{-4}$ \tabularnewline
\hline
\end{tabular}
\caption{Constraints from loop-level lepton flavour violating interactions and anomalous magnetic moments. Experimental limits are taken from~\cite{MEG:2016leq,BaBar:2009hkt}.}\label{tab:loop}
\end{table}

The interactions of the singly-charged scalar generate four-fermion operators that lead to  a charged current involving four leptons $\ell_i\to \ell_j\nu\nu$. These processes are highly constrained from  lepton universality tests. The current constraints are given by~\cite{HFLAV:2019otj},
\begin{align}
&\frac{g_\tau}{g_e}=\left| \frac{1+\delta(\tau\to \mu\nu\nu)}{1+\delta(\mu\to e\nu\nu)} \right|= 1.0029\pm 0.0014,
\\&
\frac{g_\mu}{g_e}=\left| \frac{1+\delta(\tau\to \mu\nu\nu)}{1+\delta(\tau\to e\nu\nu)} \right|= 1.0018\pm 0.0014,
\\&
\frac{g_\tau}{g_\mu}=\left| \frac{1+\delta(\tau\to e\nu\nu)}{1+\delta(\mu\to e\nu\nu)} \right|= 1.0010\pm 0.0014.
\end{align}
We impose $2\sigma$ constraints on these quantities, and  $\delta$ is defined as,
\begin{align}
\delta(\ell_i\to \ell_j\nu\nu) =|f_{ji}|^2 v^2/m^2_\phi,    
\end{align}
where as usual $v=246$ GeV. Furthermore, not to change the Fermi constant from the SM prediction, we impose $2\delta(\mu\to e\nu\nu)< 0.002$~\cite{Marciano:2000yj}. 

The singly- and the doubly-charged scalars do not interact with the quarks; therefore, lepton flavor violating $\mu - e$ conversion in the nuclei is forbidden at the tree-level. Such processes are then induced at one-loop level and are strongly correlated with $\mu\to e \gamma$. However, as long as the latter mode satisfies the current experimental bounds, the  $\mu - e$ conversion is also below the present bound. This is why we do not include the corresponding bounds in our analysis.

The current experimental bounds and the associated constraints on the model parameters arising from tree-level mediated cLFV processes are summarized in Table~\ref{tab:tree}. Similarly, the current limits on cLFV decays and lepton anomalous magnetic moment that happen at the one-loop level along with the constraints are  recapitulate in  Table~\ref{tab:loop}.

\subsection{Current collider constraints} 

At the LHC, both the singly- and the doubly-charged scalars are  pair produced via Drell-Yan mechanism. Recently, ATLAS collaboration has presented their updated search on the doubly-charged scalar in the same-charge two-lepton invariant mass spectrum. In the context of Zee-Babu model,  their study includes $e^\pm e^\pm, e^\pm \mu^\pm,$ and $\mu^\pm \mu^\pm$ final states with the identification of three/four leptons. With 139~${\rm fb}^{-1}$ of data  from proton–proton collisions at $\sqrt{s}$= 13 TeV, no significant excess above the SM prediction was found. This rules out the Zee-Babu state, $\kappa^{\pm\pm}$,  with masses smaller than 900 GeV~\cite{ATLAS:2022pbd}, whereas,  the projected sensitivity with 3~${\rm ab}^{-1}$ of data at $\sqrt{s}$= 13 TeV, the limit is 1110 GeV~\cite{Ruiz:2022sct}. However, there are no dedicated searches for the weak-singlet singly-charged scalar at the LHC. Recently, Ref.~\cite{Alcaide:2019kdr} performed an analysis of such scalars and found that a charged scalar decaying exclusively into $e^\pm$ and $\mu^\pm$ can be excluded up to masses of 500 GeV with an integrated luminosity of 200  ${\rm fb}^{-1}$ data.

\section{Numerical Analysis}
\label{sec:numerical}

In this section, we perform a numerical fit to the neutrino oscillation data. The neutrino mass matrix is given in equation~\eqref{eq:numass} that contains nine Yukawa couplings $f_{ij}, g_{ij}$ and a cubic coupling constant $\mu$. Without loss of generality, the charged lepton mass matrix can be made real, diagonal, and positive. In this basis, via field redefinition, one can make elements of $f$ real. By further field redefinition, one of the phases from $g$ can be removed as well as the cubic coupling can be made real. This leaves us with ten real parameters and five phases. However, for the simplicity of our analysis,  we consider all parameters to be real (i.e., set all phases to zero). For the masses of the charged leptons appearing in the neutrino mass formula, we use the PDG values~\cite{ParticleDataGroup:2020ssz}.  In the fitting procedure, we make sure that the parameters $\left(f_{ij}, g_{ij}, \mu\right)$ satisfy all the theoretical constraints mentioned above.  Furthermore, the loop function entering in neutrino mass formula involves masses of the two scalars, which we fix to certain values. 

This numerical study is based on a $\chi^2$ analysis, and which is defined as,
\begin{align}
\chi^2= \sum_i  \left( \frac{T_{i}-E_{i}}{\sigma_i}\right)^2.
\end{align}
Here $\sigma_i$ represents experimental uncertainty; $E_{i}$ and $T_{i}$  stand for experimental central value  and theoretical prediction for the $i$-th observable respectively. In the above equation, $i$ is summed over five observables, two neutrino mass-squared differences and three mixing angles.  Experimental values of these quantities are given in Table~\ref{tab:nuEXP}. In this fitting procedure, we do not include the CP violating phase, $\delta$, in the $\chi^2$-function. This phase has not been measured yet in the experiments, and a global fit to the neutrino oscillation data currently allows almost the entire range from 0 to $2\pi$.  

Since the parameters, for a fixed set of masses $(m_\phi, m_\kappa)$, are severely constrained by the flavor violating processes, we perform a constrained minimization. In this procedure, the  $\chi^2$ function is subject to all constraints, theoretical as well as experimental, discussed in Sec.~\ref{sec:constraints}. Since it is highly challenging to explore the entire parameter space,  we consider five benchmark points. Fit parameters obtained for the benchmark points are listed in Table~\ref{tab:BSs:BRs}. Moreover, in the same table, we summarize various branching ratios  for $\phi^\pm$ and $\kappa^{\pm\pm}$ decay modes. The outcome of these benchmark fits are presented in Table~\ref{tab:BSs:Low}. In this work, we only focus on NO for the neutrino masses, which demands the largest Yukawa coupling to be $g_{\mu\mu}$, hence has profound implication in the muon colliders. If instead, IO is considered, the largest Yukawa coupling becomes $g_{\mu\tau}$ (see, for example, Ref.~\cite{Herrero-Garcia:2014hfa}).

\begin{table*}[th!]
\setlength\tabcolsep{17pt}
\begin{center}
\begin{adjustbox}{max width=1.01\textwidth}
\begin{tabular}{lccccc}
\toprule
\multicolumn{1}{c} { Benchmark point }&  BP1 & BP2 & BP3 & BP4 & BP5 \\
\midrule
\multicolumn{6}{c}{\textit{Parameters}} \\
\midrule
$m_\kappa~({\rm GeV})$ & $1250$  &   $1250$ & $2500$ & $1250$ & 3750 \\
$m_\phi~({\rm GeV})$ & $1250$ & $2500$ & $1250$  & $3750$  & 1250 \\
$\mu~({\rm GeV})$ & $1903.01$ & $1957.01$ & $1994.75$ & $1730.09$ & 2067.06 \\ 
$f_{e\mu}$   & $-0.03809$ &   $-0.06687$ & $-0.02157$ & $-0.1026$  & -0.03558  \\
$f_{e\tau}$   & $0.02037$      &    $0.03577$   & $0.02918$ & $0.05487$  & 0.01925  \\
$f_{\mu\tau}$ & $0.06297$  &  $0.11052$ & $0.05291$ & $0.16973$  & 0.05893 \\
$g_{ee}$ & $-0.19669$   & $-0.02474$ & $-0.02499$  & $-0.01731$  & -0.00269 \\
$g_{e\mu}$ &  $9.89\times 10^{-6}$  & $-5.05\times 10^{-4}$ & $-0.00237$  & -0.00160 & 0.00132 \\
$g_{e\tau}$   & $0.00462$ & $0.00289$  & $0.04409$  &  $0.02005$  & -0.00699  \\
$g_{\mu\mu}$  & $0.48$  &  $0.487$ & $0.99$ & $0.488$ & 1.0 \\ 
$g_{\mu\tau}$  & $0.02542$ & $0.02579$ &  $0.05029$ & $0.02582$  &  0.05270\\ 
$g_{\tau\tau}$  & $0.00222$  & $0.00225$ & $0.00420$ & $0.00225$ &  0.00457 \\
\midrule
\multicolumn{6}{c}{{\it Decays of $\phi^\pm$}} \\
\midrule
${\rm BR}(\phi^+ \to \mu^+ \bar{\nu}_\tau)$   & $3.33 \times 10^{-1}$ & $3.27 \times 10^{-1}$ & $3.33 \times 10^{-1}$ & $3.23 \times 10^{-1}$  & $3.32 \times 10^{-1}$ \\
${\rm BR}(\phi^+ \to \tau^+ \bar{\nu}_\mu)$   & $3.33 \times 10^{-1}$ & $3.27 \times 10^{-1}$ & $3.33 \times 10^{-1}$ & $3.23 \times 10^{-1}$  & $3.32 \times 10^{-1}$ \\
${\rm BR}(\phi^+ \to e^+ \bar{\nu}_\mu)$ & $1.22 \times 10^{-1}$ & $1.19 \times 10^{-1}$ & $5.52 \times 10^{-2}$ & $1.18 \times 10^{-1}$  & $1.21 \times 10^{-1}$ \\
${\rm BR}(\phi^+ \to \mu^+ \bar{\nu}_e)$   & $1.22 \times 10^{-1}$ & $1.19 \times 10^{-1}$ & $5.52 \times 10^{-2}$ & $1.18 \times 10^{-1}$  & $1.21 \times 10^{-1}$ \\
${\rm BR}(\phi^+ \to e^+ \bar{\nu}_\tau)$   & $3.48 \times 10^{-2}$ & $3.42 \times 10^{-2}$ & $1.01 \times 10^{-1}$ & $3.38 \times 10^{-2}$  & $3.54 \times 10^{-2}$ \\
${\rm BR}(\phi^+ \to \tau^+ \bar{\nu}_e)$  & $3.48 \times 10^{-2}$ & $3.42 \times 10^{-2}$ & $1.01 \times 10^{-1}$ & $3.38 \times 10^{-2}$  & $3.54 \times 10^{-2}$ \\
${\rm BR}(\phi^+ \to W^+ \bar{\nu}_i \bar{\nu}_j)$  & $1.48 \times 10^{-2}$ & $2.72 \times 10^{-2}$ & $1.52 \times 10^{-2}$ & $4.97 \times 10^{-2}$ & $1.95 \times 10^{-2}$ \\ 
$\Gamma_{\phi^\pm}~({\rm GeV})$  & $1.18$ & $7.43$ & $0.84$ & $26.59$ & $1.04$ \\
\midrule
\multicolumn{6}{c}{{\it Decays of $\kappa^{\pm\pm}$}} \\
\midrule
${\rm BR}(\kappa^{++} \to \mu^+ \mu^+)$  & $8.52 \times 10^{-1}$ & $9.92 \times 10^{-1}$ & $9.75 \times 10^{-1}$ & $9.89 \times 10^{-1}$  & $8.11 \times 10^{-1}$ \\
${\rm BR}(\kappa^{++} \to e^+ e^+)$  & $1.43 \times 10^{-3}$ & $2.55 \times 10^{-4}$ & $6.21 \times 10^{-4}$ & $1.24 \times 10^{-3}$  & $5.87 \times 10^{-6}$ \\
${\rm BR}(\kappa^{++} \to \mu^+ \tau^+)$ & $4.78 \times 10^{-3}$ & $5.56 \times 10^{-3}$ & $5.03 \times 10^{-3}$ & $5.97 \times 10^{-3}$  & $4.51 \times 10^{-3}$ \\
${\rm BR}(\kappa^{++} \to e^+ \tau^+)$  & $1.57 \times 10^{-4}$ & $6.98 \times 10^{-5}$ & $3.87 \times 10^{-3}$ & $3.34 \times 10^{-3}$  & $7.93 \times 10^{-5}$ \\
${\rm BR}(\kappa^{++} \to \tau^+ \tau^+)$  & $1.82 \times 10^{-5}$ & $2.11 \times 10^{-5}$ & $1.75 \times 10^{-5}$ & $2.10 \times 10^{-5}$  & $1.69 \times 10^{-5}$ \\
${\rm BR}(\kappa^{++} \to e^+ \mu^+)$  & $6.80 \times 10^{-10}$ & $2.13 \times 10^{-6}$ & $1.12 \times 10^{-5}$ & $2.12 \times 10^{-5}$  &  $2.82 \times 10^{-6}$\\
${\rm BR}(\kappa^{++} \to \phi^+ \phi^+)$  & $0$ & $0$ & $0$ & $0$  & $1.84 \times 10^{-1}$ \\
$\Gamma_{\kappa^{\pm\pm}}~({\rm GeV})$ & $13.45$ & $11.89$ & $99.95$ & $11.97$ & $183.84$ \\
\bottomrule
\end{tabular}
\hspace{0.2cm}
\end{adjustbox}
\end{center}
\caption{The benchmark points used in this study. Here we show also some of its characteristics like the decay branching ratios of the singly-charged ($\phi^\pm$) and doubly-charged ($\kappa^{\pm\pm}$) scalars. These benchmark parameters along with the neutrino mass matrix can be provided as a text file upon request.}
\label{tab:BSs:BRs}
\end{table*} 

\begin{table*}[th!]
\setlength\tabcolsep{13pt}
\begin{center}
\begin{adjustbox}{max width=1.01\textwidth}
\begin{tabular}{lccccc}
\toprule
\multicolumn{1}{c} { Benchmark point } & BP1 & BP2 & BP3 & BP4 & BP5 \\
\midrule
\multicolumn{6}{c}{{\it Neutrino-oscillation data}} \\
\midrule
$m_1~({\rm eV})$ &  0 & 0 & 0 & 0 & 0\\
$m_2~({\rm eV})$ & 0.0086 &0.0086 &0.0086 &0.0086 &0.0086\\
$m_3~({\rm eV})$  & 0.0501 &0.0501 &0.0501 &0.0501 &0.0501\\
$\Delta m^2_{21}/10^{-5}~({\rm eV}^2)$  & 7.425 &7.425 &7.426 &7.425 &7.366\\
$\Delta m^2_{31}/10^{-3}~({\rm eV}^2)$ & 2.515& 2.515 & 2.514 & 2.511& 2.511\\
$\sin^2\theta_{12}$  & 0.3045 & 0.3046 & 0.3045 & 0.3044 & 0.3049\\
$\sin^2\theta_{13}$  & 0.02223 & 0.02222 &0.02223 & 0.02223 &0.02217\\ 
$\sin^2\theta_{23}$ & 0.572 & 0.572 & 0.573 & 0.573 & 0.568\\
$m_{\beta\beta}\; (meV)$  & 3.68 & 3.68 & 1.45 & 3.67 & 3.67\\
\midrule
\multicolumn{6}{c}{{${\rm BR}(\ell_i \to \ell_j \gamma)$}}\\
\midrule
${\rm BR}(\mu \to e\gamma)$   &\bm{$2.93\times 10^{-13}$}  &\bm{$2.15\times 10^{-13}$}  &\bm{$3.73\times 10^{-13}$}  &\bm{$3.06\times 10^{-13}$ } &\bm{$2.29\times 10^{-13}$} \\
${\rm BR}(\tau \to e\gamma)$ &$5.19 \times 10^{-13}$ &$9.77 \times 10^{-14}$ &$6.62 \times 10^{-14}$ &$1.56 \times 10^{-13}$ &$1.22 \times 10^{-13}$\\
${\rm BR}(\tau \to \mu\gamma)$   &$6.51 \times 10^{-11}$ &$6.90 \times 10^{-11}$ &$6.74 \times 10^{-11}$ &$6.91 \times 10^{-11}$ &$1.50 \times 10^{-11}$\\
\midrule
\multicolumn{6}{c}{{${\rm BR}(\ell_i \to \ell_j \ell_k \ell_k)$}}\\
\midrule
${\rm BR}(\mu^{-}\to e^{+}e^{-}e^{-})$   &\bm{$2.85\times 10^{-15}$} &\bm{ $1.17\times 10^{-13}$}  &\bm{ $1.65\times 10^{-13}$}   &\bm{ $5.78\times 10^{-13}$}   &\bm{ $1.18\times 10^{-16}$} \\
${\rm BR}(\tau^{-}\to e^{+}e^{-}e^{-})$  &$1.11\times 10^{-10}$ &$6.88\times 10^{-13}$ &$1.02\times 10^{-11}$ &$1.62\times 10^{-11}$ &$5.87\times 10^{-16}$\\
${\rm BR}(\tau^{-}\to e^{+}e^{-}\mu^{-})$   &$5.06\times 10^{-19}$  &$5.74\times 10^{-16}$ &$1.84\times 10^{-13}$ &$2.67\times 10^{-13}$ &$2.84\times 10^{-16}$\\
${\rm BR}(\tau^{-}\to e^{+}\mu^{-}\mu^{-})$ &$6.59\times 10^{-10}$  &$2.66\times 10^{-10}$  &\bm{$1.59\times 10^{-8}$}  &\bm{$1.28\times 10^{-8}$}  &$8.11\times 10^{-11}$\\
${\rm BR}(\tau^{-}\to \mu^{+}e^{-}e^{-})$ &\bm{$3.26\times 10^{-9}$} &$5.32\times 10^{-11}$ &$1.29\times 10^{-11}$ &$2.61\times 10^{-11}$ &$3.24\times 10^{-14}$\\
${\rm BR}(\tau^{-}\to \mu^{+}e^{-}\mu^{-})$ &$1.65\times 10^{-17}$ &$4.44\times 10^{-14}$ &$2.32\times 10^{-13}$ &$4.46\times 10^{-13}$ &$1.57\times 10^{-14}$\\
${\rm BR}(\tau^{-} \to \mu^{+}\mu^{-}\mu^{-})$   &\bm{$1.94\times 10^{-8}$}  &\bm{$2.06\times 10^{-8}$}   &\bm{$2.02\times 10^{-8}$}   &\bm{$2.07\times 10^{-8}$}  &\bm{$4.48\times 10^{-9}$} \\
\bottomrule
\end{tabular}
\hspace{0.2cm}
\end{adjustbox}
\end{center}
\caption{Predictions of the model of the neutrino masses and mixings, and lepton-flavor violating decay branching ratios. The neutrinoless double beta decay parameter is defined as $m_{\beta\beta}\equiv \left|\sum_i U^2_{ei}m_i\right|$.  Predictions of the cLFV modes that are just below the current bound and  will be fully tested in the next upgrades are highlighted with boldface.} 
\label{tab:BSs:Low}
\end{table*} 

From our numerical analysis, we find that the most stringent constraint on the Yukawa parameters and the masses of the Zee-Babu states arise from the charged lepton  flavor violating $\mu\to e\gamma$ process. In particular, our result shows that without going to a fine-tuned part of the parameter space, the minimum masses possible for the Zee-Babu states is close to $\sim$900 GeV, which is fully consistent with the previous findings~\cite{Schmidt:2014zoa,Herrero-Garcia:2014hfa}. Fine-tuned regions of the parameter space do allow Zee-Babu states to have masses as low as $\sim \mathcal{O}(200)$ GeV~\cite{Herrero-Garcia:2014hfa}.  It is important to note that even if other important flavor violating processes, for example, $\mu\to 3e$, may be suppressed by taking vanishing $g_{ee}$ or $g_{e\mu}$, however, $\mu\to e\gamma$ cannot be suppressed to arbitrarily small values since both the Yukawa couplings $f$ and $g$ contribute to it that cannot taken to be zero simultaneously.  We emphasise that it is still difficult to fully rule out the Zee-Babu model from cLFV constraints. Moreover, the BSM scalars can be as heavy as $\sim \mathcal{O}(100)$ TeV depending on perturbativity conditions imposed; for a detailed discussion on these, we refer the readers to Refs.~\cite{Schmidt:2014zoa,Herrero-Garcia:2014hfa,Babu:2015ajp}.

A fit to the neutrino oscillation data requires the largest entry to be the $\mu\mu$ coupling with the doubly-charged scalar.  The coupling of the singlet is also expected to be sizable with the muon.  This is why, muon colliders are the perfect machines to search for these states that have masses in the multi TeV range. In our studied five benchmark scenarios, we have fixed the masses of these states to be $(m_\phi, m_\kappa)=$ $(m,m)$, $(2 m,m)$, $(m,2 m)$, $(3 m,m)$,  and $(m,3 m)$ (with $m=1250$ GeV), which have great potential to be discovered in the muon colliders.  All these benchmark scenarios also predict large flavor violating processes in the muon and tau decays, specifically $\mu\to e\gamma$, $\mu\to 3e$ and $\tau\to 3\mu$,   $\tau\to e^+\mu\overline\mu$ modes are lying just below the current limit and will be observed in the near future,  see Table~\ref{tab:BSs:Low}. The future sensitivities of these flavor violating processes are BR$\sim 6\times 10^{-14}$ for $\mu\to e\gamma$~\cite{Baldini:2013ke},  BR$\sim 10^{-16}$ for  $\mu\to 3e$~\cite{Blondel:2013ia}, and BR$\sim 10^{-9}$ for $\tau\to 3\mu$,   $\tau\to e^+\mu\overline\mu$~\cite{Aushev:2010bq}.  These promising flavor violating processes that are within reach of near future experiments are highlighted with boldface in  Table~\ref{tab:BSs:Low}. 

Before concluding this section, we remark that another interesting neutrino mass model, namely, the type II seesaw that leads to neutrino mass at tree-level, also predicts a singly- and a doubly-charged scalars arising from a scalar weak-triplet. Phenomenology of this model is different from that of Zee-Babu model and there are much more freedom in choosing the Yukawa couplings to fit oscillation data. A comprehensive study of type II seesaw model and distinguishing between these two models at muon colliders will be presented in a future work.

\section{The Zee-Babu model at muon colliders: production rates}
\label{sec:collider}
In this section, we analyse the production rates for the Zee-Babu states at multi--TeV muon colliders. The different production channels can be categorised into three classes: ({\it i}) the production of charged lepton pairs, ({\it ii}) the production of singly-charged scalars, and ({\it iii}) the production of doubly-charged scalars. For all the calculations of the production rates, we make use of \textsc{Madgraph5\_aMC@NLO} version 3.4.2 \cite{Alwall:2014hca} with the use of the publicly available \textsc{FeynRules} \cite{Alloul:2013bka} model file which can be found in this link \url{https://feynrules.irmp.ucl.ac.be/wiki/ZeeBabu}.  
\subsection{Charged lepton pair production}

\begin{figure}
    \centering
    \includegraphics[width=0.65\linewidth]{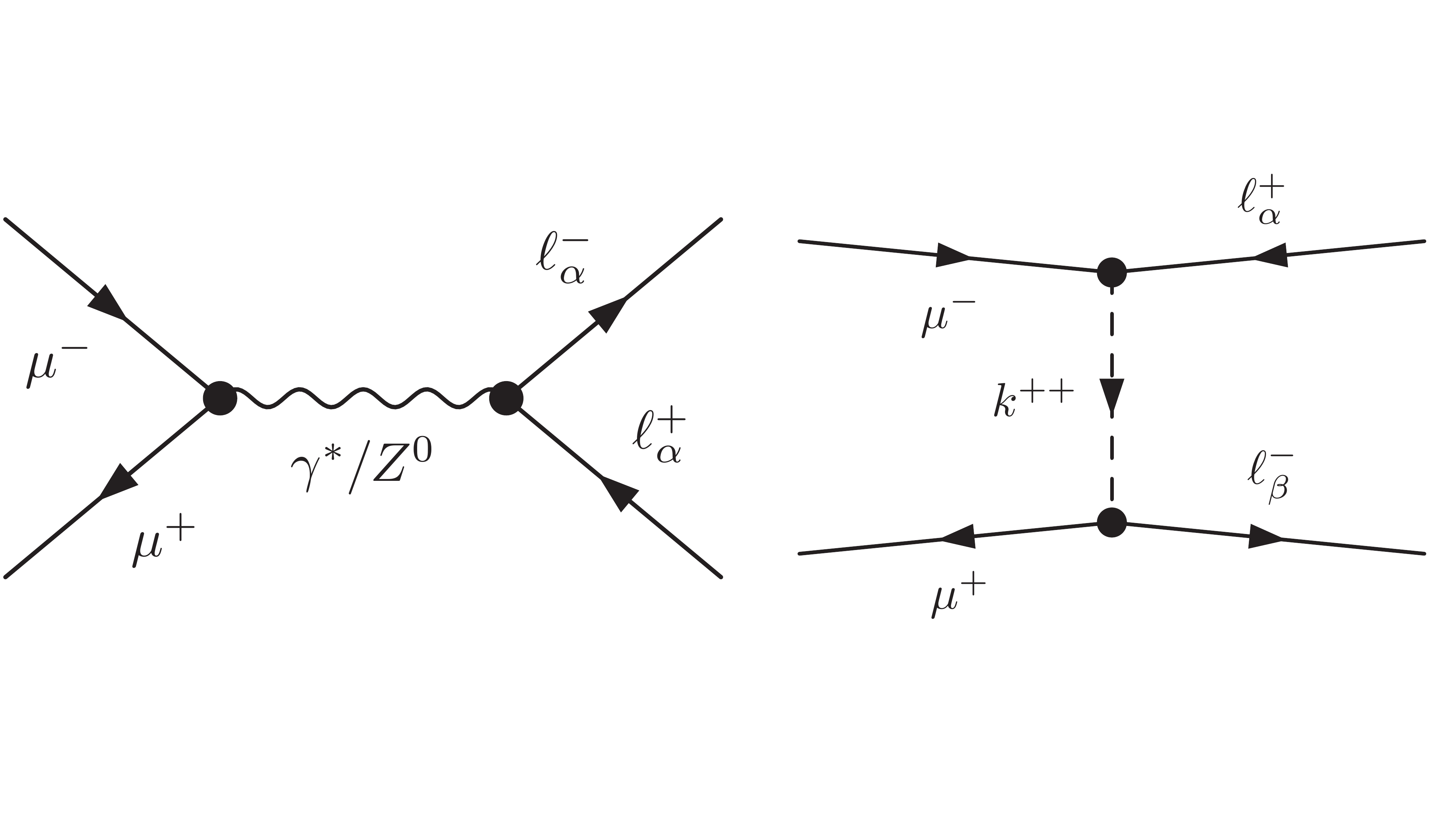}
    \vspace{-1cm}
    \caption{Examples of Feynman diagrams for the pair production of charged leptons at muon colliders. The Drell-Yan type diagram on the left contributes only for the case of flavour-conserving case while the $t$--channel diagram on the right contributes to both the flavour conserving ($\alpha=\beta$) and the flavour violating ($\alpha \neq \beta$) production channels.}
    \label{fig:fd:dilepton}
\end{figure}

\begin{figure}[!t]
    \centering
    \includegraphics[width=0.495\linewidth]{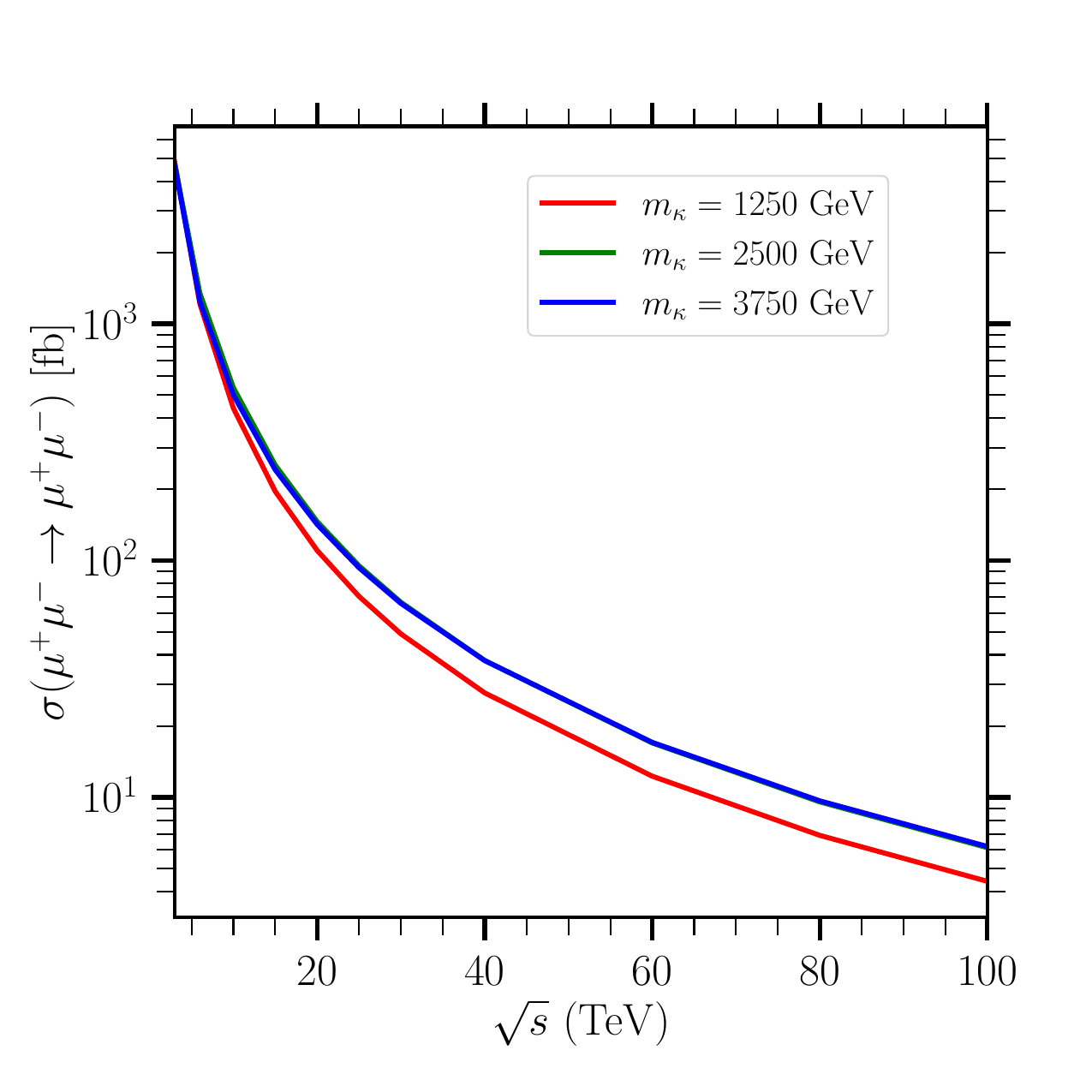}
    \hfill 
    \includegraphics[width=0.495\linewidth]{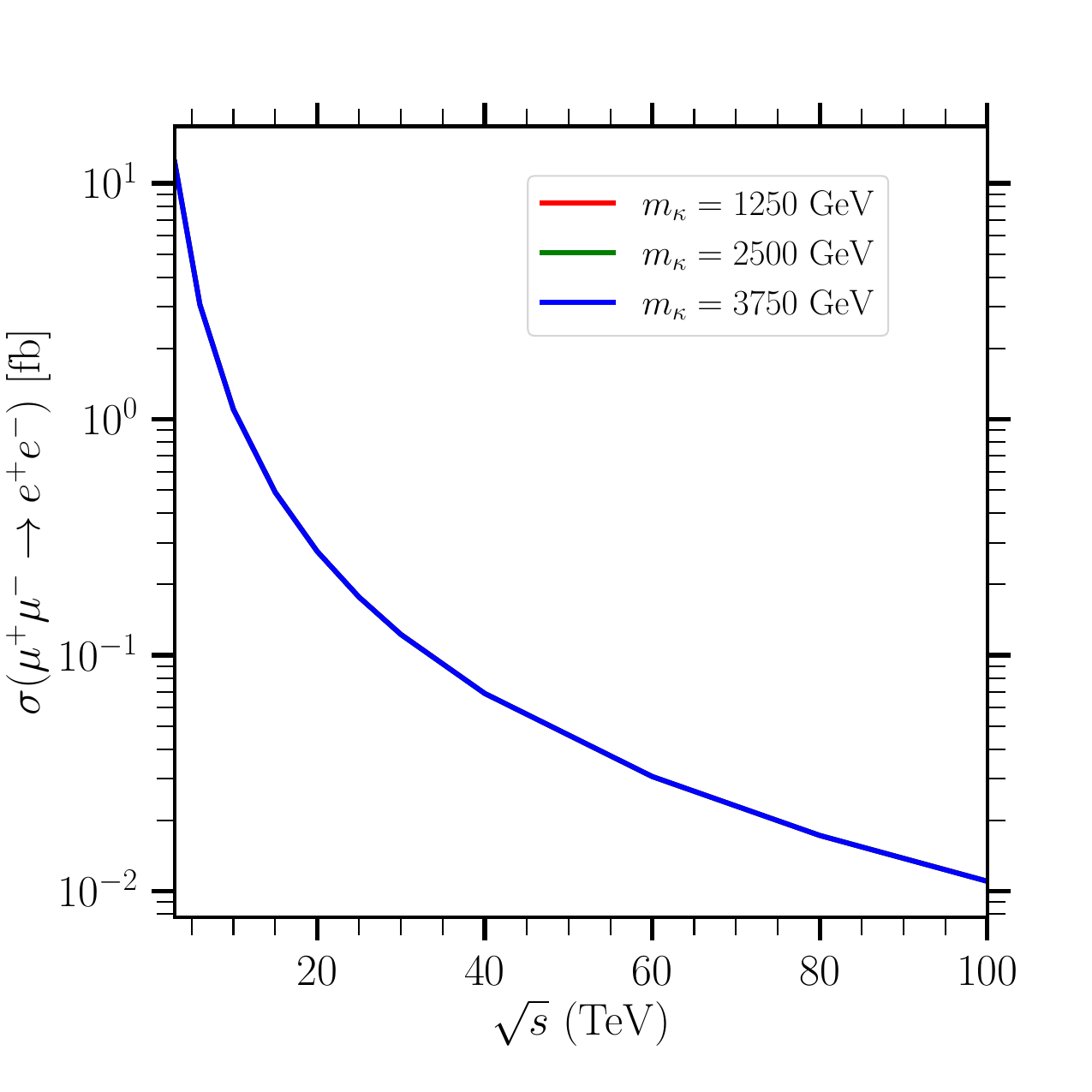}
    \caption{Production cross section in femtobarn of FC lepton pair production as function of the center-of-mass energy for the muon channel (left) and electron channel (right). The results are shown for $m_\kappa=1250$ GeV (red), $m_\kappa = 2500$ GeV (green) and $m_\kappa = 3750$ GeV (blue). 
The results for $m_\kappa = 2500$ GeV and $m_\kappa = 3750$ GeV for $\mu\mu$ production on the left panel are equal. We provide more details about this on the text.}
    \label{fig:xs:FC:lepton}
\end{figure}

\begin{figure}[!t]
    \centering
    \includegraphics[width=0.495\linewidth]{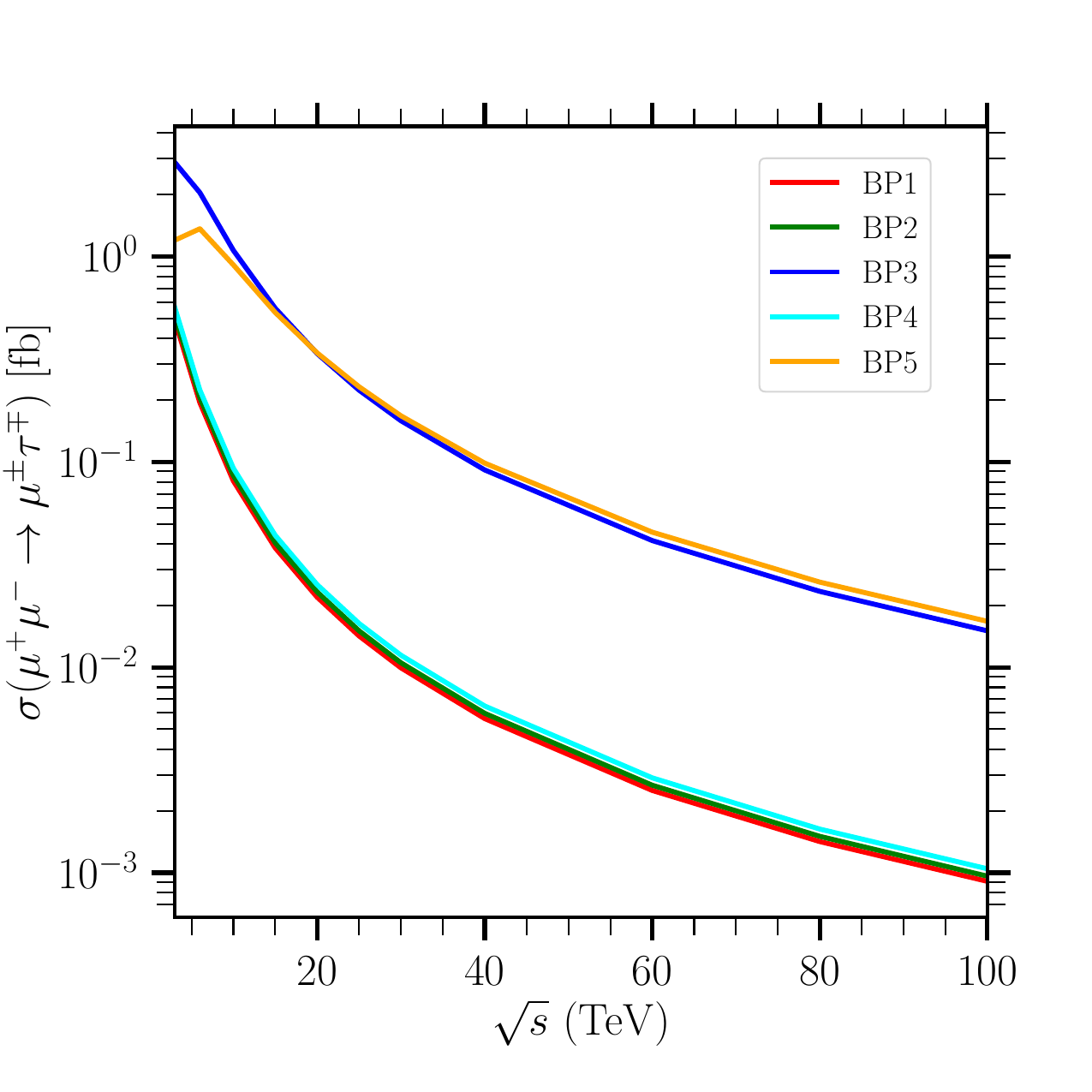}
    \hfill 
    \includegraphics[width=0.495\linewidth]{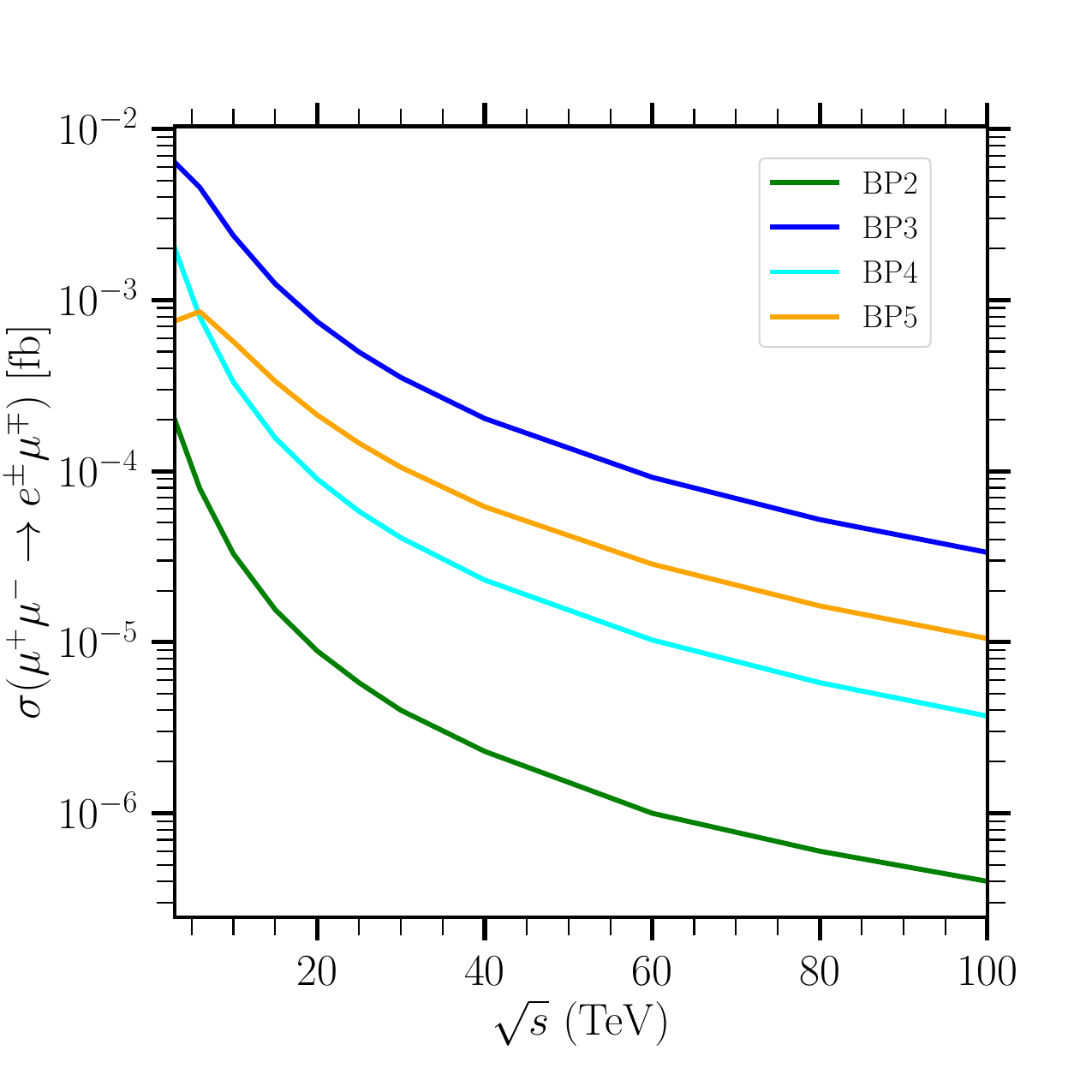}
    \caption{Production cross section in femtobarn of FV lepton pair production as function of the center-of-mass energy for the $\mu^\pm \tau^\mp$ channel (left) and $e^\pm \mu^\mp$ channel (right). The results are shown for BP1 (red), BP2 (green), BP3 (blue), BP4 (cyan) and BP5 (orange). Note that for the production of the $e^\pm \mu^\mp$, the cross section in BP1 is not shown as it is well below $10^{-8}$ fb.}
    \label{fig:xs:FV:lepton}
\end{figure}

For charged lepton pair production, we consider both the flavour-conserving (FC) as well as the flavour-violating (FV) channels (see figure \ref{fig:fd:dilepton}). The FC lepton pair production cross section is shown in figure \ref{fig:xs:FC:lepton}. The contribution of the doubly-charged scalar to the production cross section for the lepton flavour $\ell$ is proportional to $g_{\ell\ell}^2$. As per the benchmark points we have chosen a noticeable effect can be seen in the muon channel since the $g_{\kappa\mu\mu}$ coupling can be of order ${\cal O}(1)$. While a suppression of the muonic cross section is caused by the $\propto m_\ell/m_{\kappa}^2$ factor. Such effects can be seen clearly in the right panel of figure \ref{fig:xs:FC:lepton}. We can see that the effects of the doubly-charged scalars are mainly driven by the value of the $g_{\mu\mu}$ coupling since we can see that the predictions for $m_\kappa=2.5$ TeV (green) and $m_\kappa=3.75$ TeV (blue) are exactly the same because they correspond to the benchmark points BP3 and BP5 for which $g_{\mu\mu} \approx 1$. For the case of the electron channel, the contribution of the doubly-charged scalar is negligibly small since the corresponding coupling is of order ${\cal O}(10^{-2}$--$10^{-1}$) and therefore the value of the corresponding cross section is mainly driven by the contribution of the gauge-boson Feynman diagrams. The contribution to the $\tau$--lepton pair production is even much more smaller for which reason we do not show the corresponding result in  this paper. The cross section for the FV lepton pair production is shown in figure \ref{fig:xs:FV:lepton}. Contrarily to the FC case, the cross section in this case receives only contribution from the $t$--channel diagrams and has the following behavior:
\begin{equation}
    \sigma_{e^\pm \mu^\mp}~:~\sigma_{e^\pm \tau^\mp}~:~\sigma_{\mu^\pm \tau^\mp} ~\propto~ g_{\mu\mu}^2 g_{e \mu}^2~:~g_{e\mu}^2 g_{\mu\tau}^2~:~g_{\mu\mu}^2 g_{\mu\tau}^2.
\end{equation}

From the choice of the couplings $g_{ij}$ in our benchmark scenarios, it is expected to have $\sigma_{\mu^\pm \tau^\mp} \gg \sigma_{e^\pm \mu^\mp} \gg \sigma_{e^\pm \tau^\mp}$ a feature that can be clearly seen in figure \ref{fig:xs:FV:lepton}. We note that the cross section for the $e^\pm \tau^\mp$ channel is orders of magnitudes smaller and therefore we dot not show it here. We note that these processes are strongly correlated to charged lepton flavour violating decays. However, for our benchmark scenarios the most important probe would be the search for the production of $\mu^\mp \tau^\mp$ given that the other two are suppressed and therefore we do not expect them to give additional information.

\subsection{Production of singly-charged scalars}

\begin{figure}[!t]
    \centering
    \includegraphics[width=0.9\linewidth]{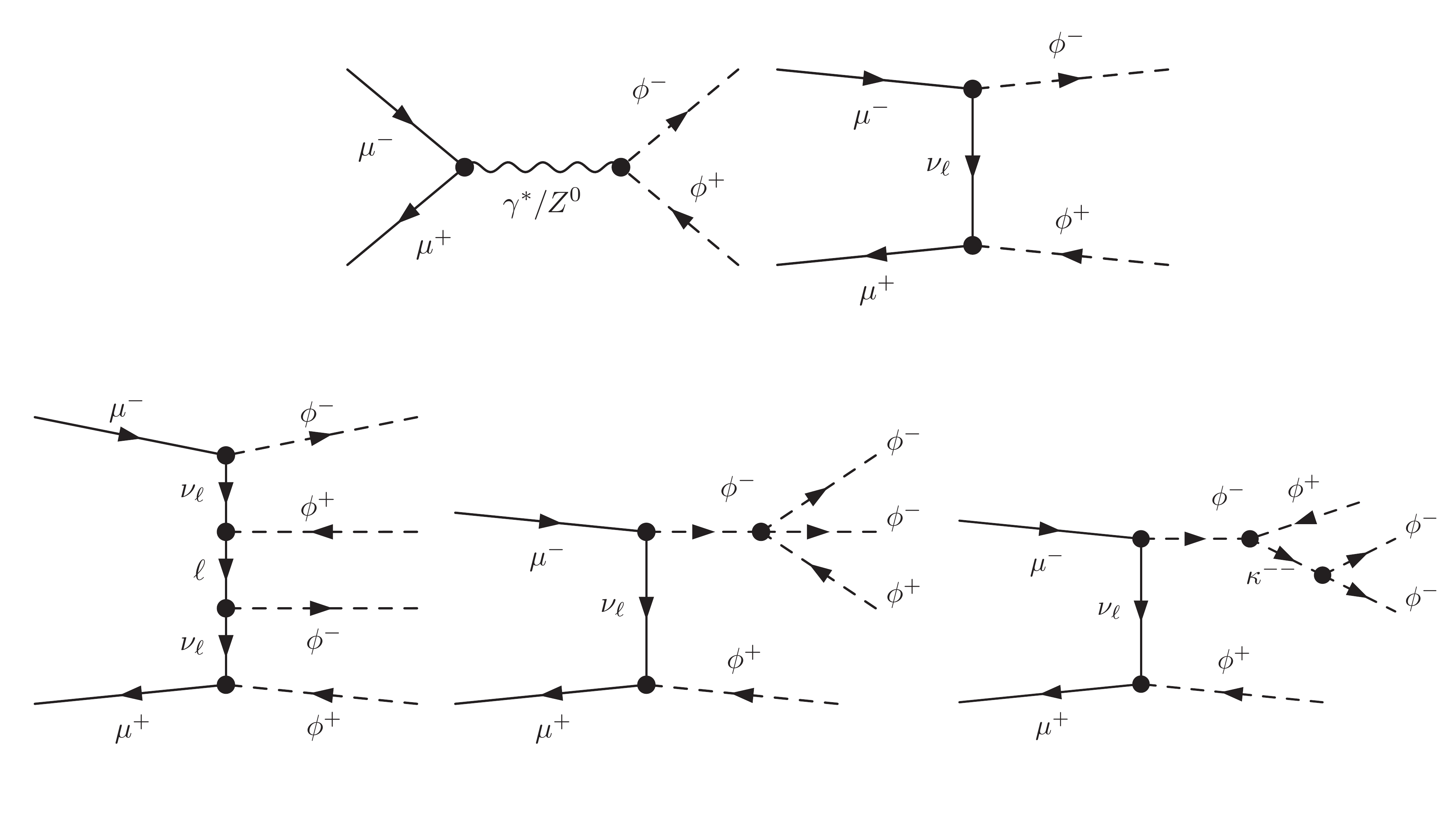}
    \caption{Example of Feynman diagrams for the production of $\phi^+ \phi^-$ (upper panel) and of $\phi^+ \phi^+ \phi^- \phi^-$ (lower panel).}
    \label{fig:fd:ss}
\end{figure}

\begin{figure}[!t]
    \centering
    \includegraphics[width=0.495\linewidth]{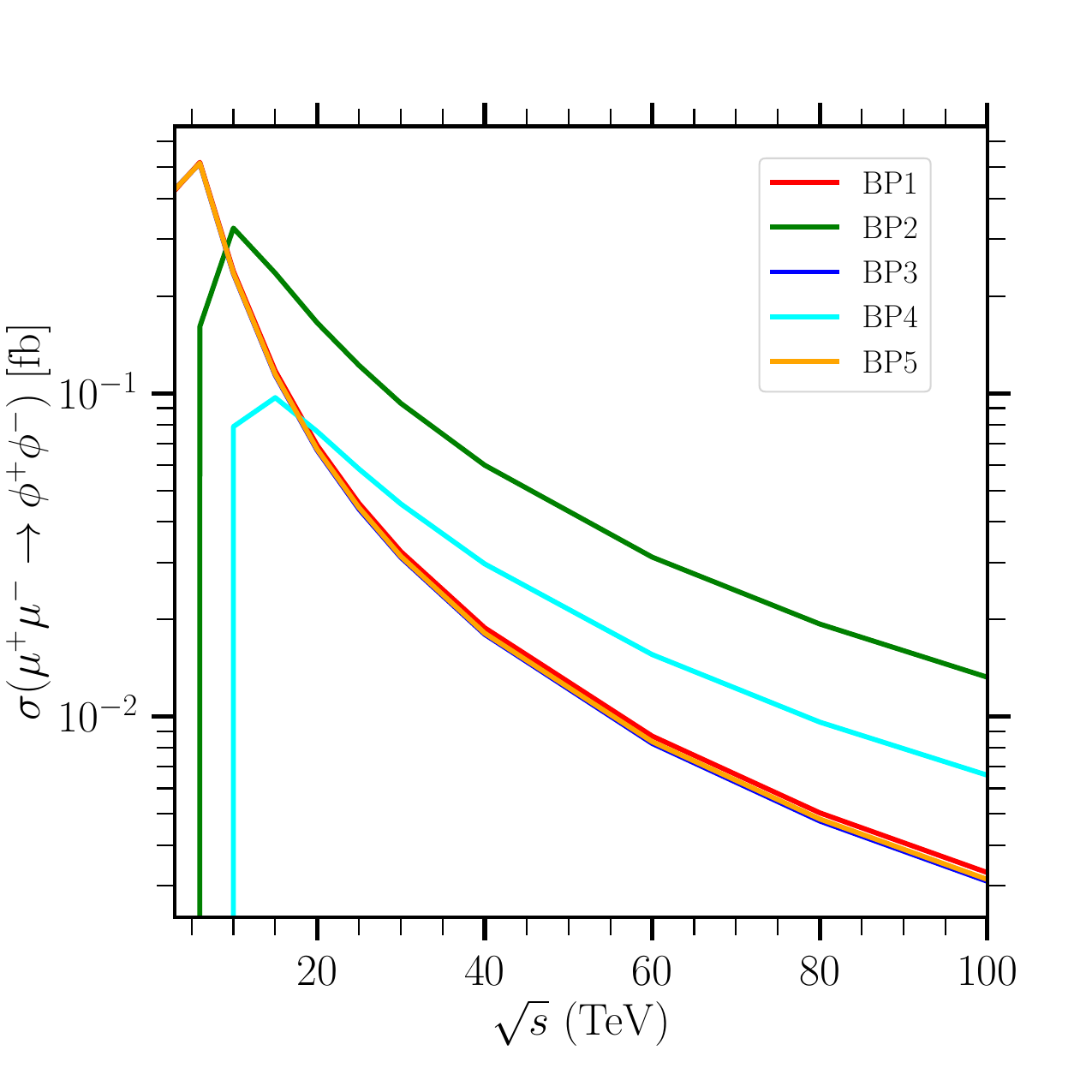}
    \hfill 
    \includegraphics[width=0.495\linewidth]{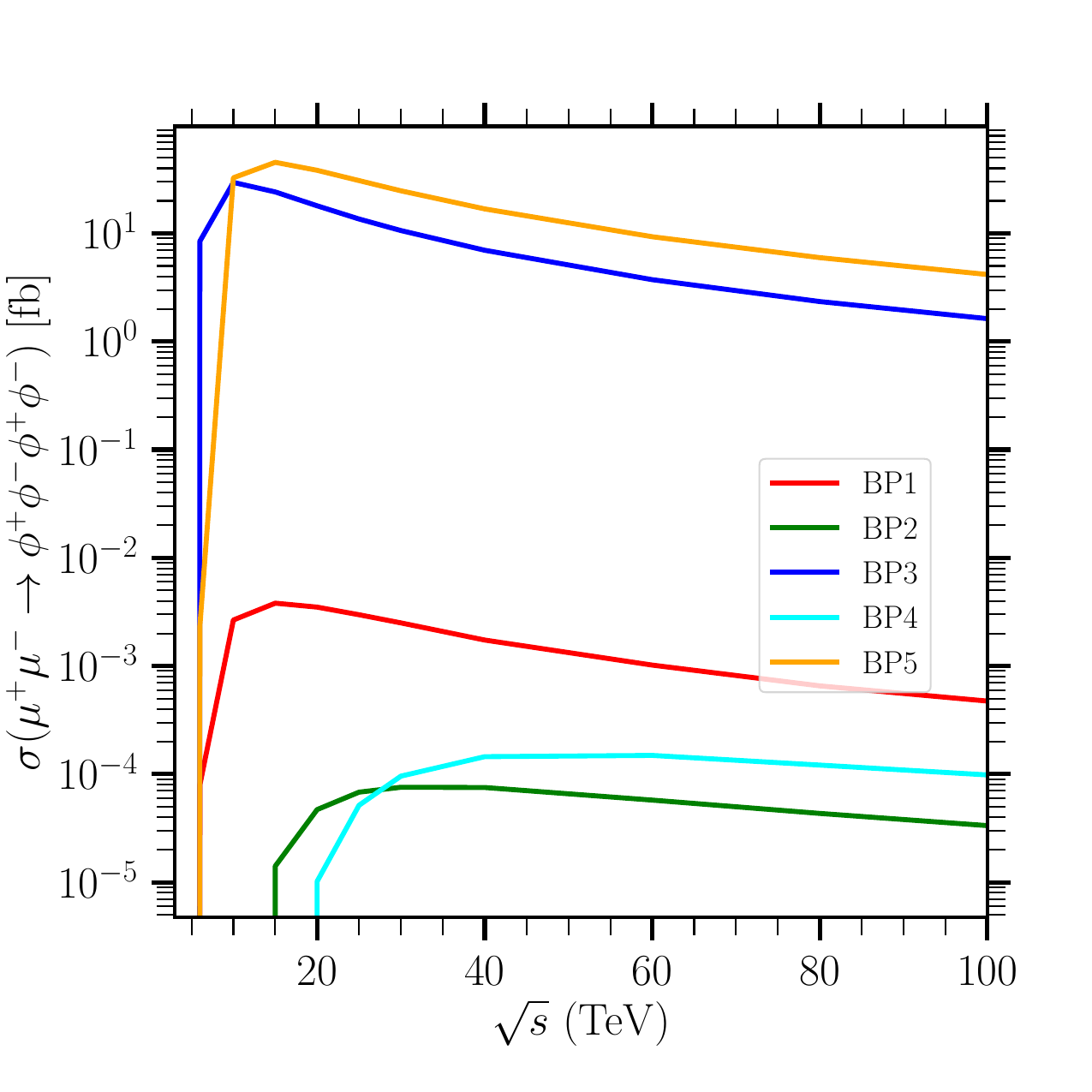}
    \caption{Same as in figure \ref{fig:xs:FV:lepton} but for the production of $\phi^+\phi^-$ (left) and of $\phi^+ \phi^- \phi^+ \phi^-$ (right).}
    \label{fig:xs:phi}
\end{figure}

We turn now into the discussion of the production mechanisms of the singly-charged scalars at muon colliders. We analyse two production channels: ({\it i}) the pair production mode (upper panel of figure \ref{fig:fd:ss}) and ({\it ii}) the production of four scalars (lower panel of figure \ref{fig:fd:ss}). The production rate for singly-charged scalar pairs receive two contributions: a Drell-Yan type contribution (left upper panel of figure \ref{fig:fd:ss}) and $t$--channel contribution (right upper panel of figure \ref{fig:fd:ss}). We note that the second contribution is always suppressed as was checked out since the relevant couplings -- $f_{e\mu}$ and $f_{\tau\mu}$ -- are small in all the benchmark scenarios under consideration. The production cross section for $\phi^+\phi^-$ fall as $1/s$ ($\sqrt{s}$ is the center-of-mass energy) as can clearly be seen in the left panel of figure \ref{fig:xs:phi} with the maximum being about $0.5$ fb for the benchmark points corresponding to $m_\phi = 1.25$ TeV (BP1, BP3 and BP5). We can see also that the total cross section for this mass value does not depend on the values of the Yukawa-type couplings which is another test of the negligible contribution of the $t$-channel diagrams in this process. For the other benchmark points, the production cross section reaches a maximum just above the threshold and then decreases as $1/s$ with the cross section for BP2 ($m_\phi=2.5$ TeV) is always larger than for BP4 ($m_\phi=3.75$ TeV). \\

For the production of four singly-charged scalars, there are multiple contributions that can be categorised into four classes: ({\it i}) the gauge-boson contribution through the exchange of a photon or a $Z$--boson and which occurs always in $s$--channel, ({\it ii}) the contribution of charged leptons and neutrinos through $t$--channel which is shown on the left lower panel of figure \ref{fig:fd:ss}, ({\it iii}) the contribution of the SM Higgs boson with a neutrino exchanged in the $t$-channel and ({\it iv}) the contribution of a doubly-charged scalar. We note that the contribution of the leptons is always subleading since it involves at three propagators and a factor of $f_{\mu \ell_\alpha}^2 f_{\mu \ell_\beta}^2 f_{\ell_\alpha \ell_\beta}^2$ which implies a suppression of about ${\cal O}(10^{-8}$--$10^{-12}$). The second contribution of the SM Higgs is proportional to the scalar quartic coupling $\lambda_{\phi}$ which we have chosen to be equal to one. The last contribution is the most dominant as it is proportional to $\mu^4$ and which has been chosen to be large in this analysis. Note that in this case, the cross section get threshold enhancement in scenarios where $m_\kappa \geq 2~m_\phi$ which is the case for BP3 and BP5. We can see clearly these features in the right panel of figure \ref{fig:xs:phi} as the cross section for these benchmark points can reach up to $30$ fb near the production threshold. For BP1 where $m_\kappa=m_\phi$ the cross section is about three orders of magnitude smaller. For heavy singly-charged scalars (BP2 and BP4) the cross section do not go above $\approx 10^{-4}$ fb.

\begin{figure}[!t]
    \centering
    \includegraphics[width=0.9\linewidth]{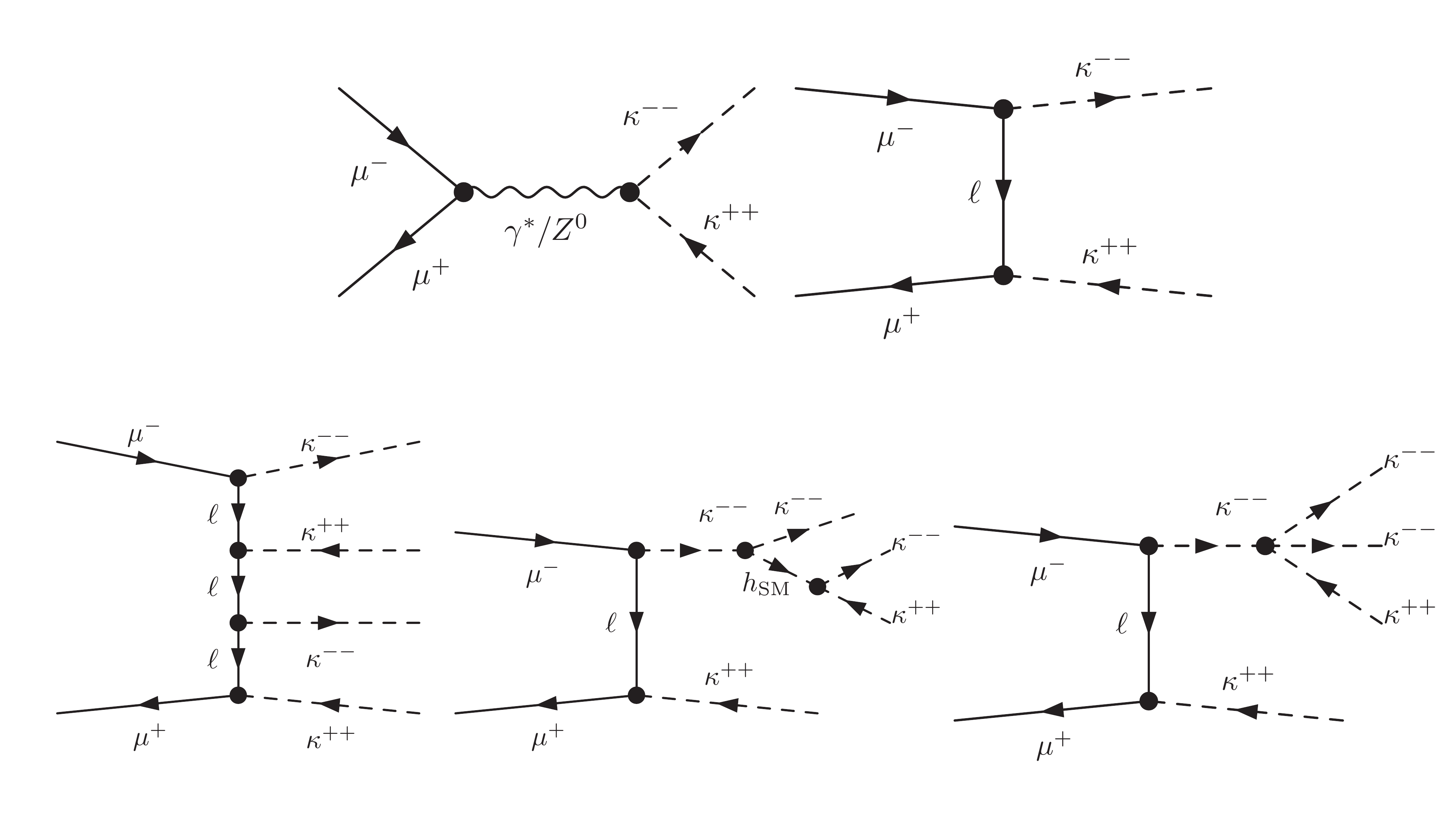}
    \caption{Example of Feynman diagrams for the production of $\kappa^{++} \kappa^{--}$ (upper panel) and of $\kappa^{++} \kappa^{++} \kappa^{--} \kappa^{--}$ (lower panel). }
    \label{fig:fd:kk}
\end{figure}

\subsection{Production of doubly-charged scalars}

\begin{figure}[!t]
    \centering
    \includegraphics[width=0.495\linewidth]{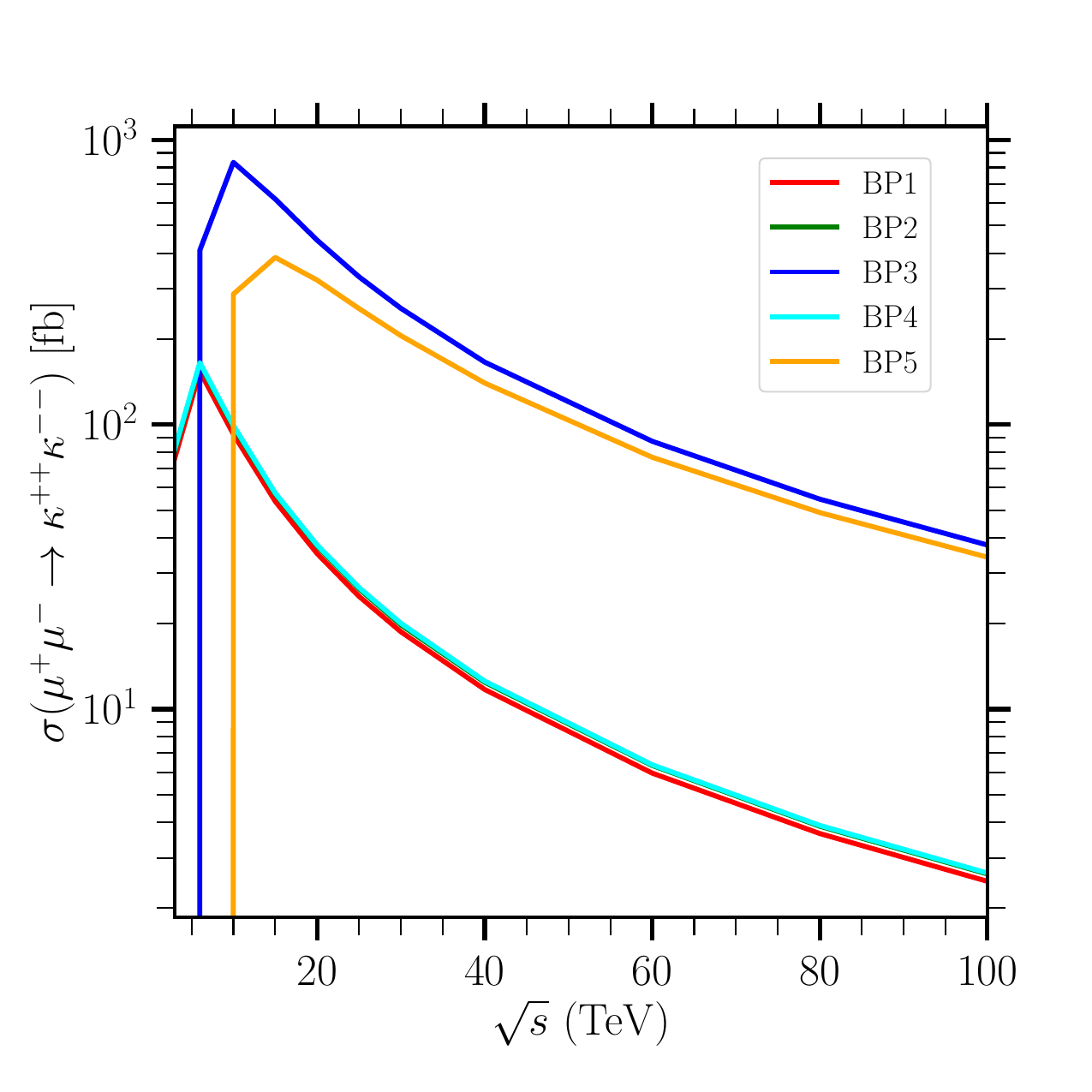}
    \hfill 
    \includegraphics[width=0.495\linewidth]{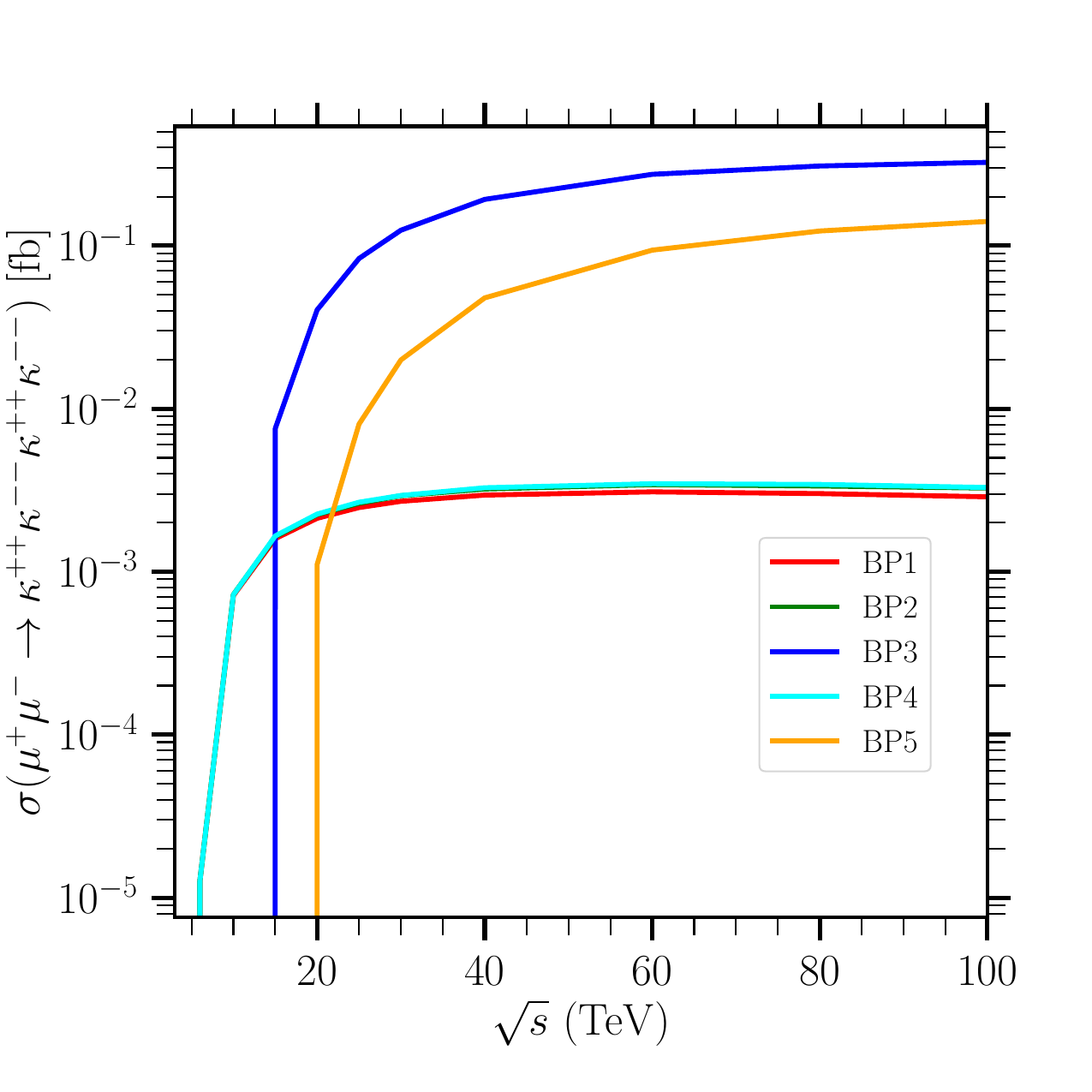}
    \caption{Same as in figure \ref{fig:xs:FV:lepton} but for the production of $\kappa^{++} \kappa^{--}$ (left) and of $\kappa^{++} \kappa^{--} \kappa^{++} \kappa^{--}$ (right). }
    \label{fig:xs:kappa}
\end{figure}

We now analyse the production of doubly-charged scalars at muon colliders. Similarly to the production of singly-charged scalars we study both the production of two scalars and of four scalars. Examples of the Feynman diagrams are depicted in figure \ref{fig:fd:kk}. The pair production of doubly-charged scalars proceeds through the contribution of Drell-Yan $s$--channel diagrams (left upper panel of figure \ref{fig:fd:kk}) and through $t$--channel diagram with the exchange of charged leptons (right upper panel of figure \ref{fig:fd:kk}). Unlike the singly-charged case, the contribution of the $t$-channel diagram to the total rate of doubly-charged pair production is very important thanks to the magnitude of the $g_{\mu\mu}$ coupling. In fact, we have checked that this contribution is the most important for all of the benchmark points; for example, it is about 20 times larger than the $s$--channel Drell-Yan for BP1 at $\sqrt{s}=3$ TeV and becomes much larger for higher energies. On the other hand, the $t$--channel contribution is found to interfere destructively with the $s$--channel contribution ($I \equiv 2 {\rm Re}({\cal M}_{\rm t}^* {\cal M}_{\rm DY})/|M_{\rm total}|^2 \approx -0.24$ at $\sqrt{s}=3$ TeV for BP1). %
In the left panel of figure \ref{fig:xs:kappa} we show the rate of the pair production cross section of doubly-charged scalars as function of $\sqrt{s}$. We can see that the production cross section in this case is more important for BP3 and BP5 for which case $g_{\mu\mu}$ is equal to $1$. It falls with the center-of-mass energy but with a minimum of about $\approx 2$ fb for BP4 which is a clear sign of the effect of the $t$--channel diagrams.

We close this section by a brief discussion of the production rate of four doubly-charged scalars at muon colliders. The Feynman diagrams are displayed on the lower panel of figure \ref{fig:fd:kk} and the production cross section is shown on the right panel of figure \ref{fig:xs:kappa}. In addition to the gauge-boson contribution, this process receives similar contribution as in the case of the singly-charged scalars except that there is no contribution from the $\mu$--term. This would imply that the total cross section is smaller than in the case of four singly-charged scalars. We can see from figure \ref{fig:xs:kappa} that the corresponding cross section varies in the range of $10^{-3}$--$10^{-1}$ fb. Two important remarks are in order here. First it seems from the same figure that the cross section violates unitarity but we have checked that basically for energies above $\sqrt{s}\approx 200$ TeV the cross section starts to decrease. Finally, despite the smallness of the corresponding cross section this channel leads to a final state of eight highly-energetic muons, which can be considered as golden channel for discovering the doubly-charged scalar at muon colliders.

\section{Sensitivity reach}
\label{ref:sensitivity}

In this section we discuss the prospects of detecting Zee-Babu states at multi--TeV muon colliders. We will perform a simple estimates of the signal-to-background ratios at both the parton and the detector levels. For this task, we will only study two channels: the pair production of the singly-charged scalars and the doubly-charged scalars. In this study we consider the sensitivity reach for $\sqrt{s}=10$ TeV and ${\cal L} =10$ fb$^{-1}$ assuming that the integrated luminosity increases linearly with the center-of-mass energy to compensate for the $1/s$ decrease in the cross section of the most processes that we can encounter at future muon colliders.

\subsection{Technical setup}
\label{sec:technical}

We consider the sensitivity reach of the Zee-Babu model at future muon colliders in the $\phi^+\phi^-$ and $\kappa^{++} \kappa^{--}$ channels. For each case, we consider first a simple analysis at the parton level by comparing the signal and the background cross sections as a function of the minimum cut on the lepton transverse momentum. To get realistic estimates of the expected sensitivity reach, we also perform a fully-fledged signal-to-background optimization taking into account all the parton-shower and detector effects. In the detector-level part of the study, we consider only the muonic decay channels, {\it i.e.}
\begin{eqnarray}
    \mu^+ \qquad \mu^- \quad &\to& \quad \phi^+ (\to \mu^+ \nu_\ell) \qquad \phi^- (\to \mu^- \bar{\nu}_\ell), \nonumber \\
    \mu^+ \qquad \mu^- \quad &\to& \quad \kappa^{++} (\to \mu^+ \mu^+) \qquad \kappa^{--} (\to \mu^- \mu^-).
\end{eqnarray}
In this case, we have signatures consisting of two muons and missing energy for the $\phi^+ \phi^-$ channel and four muons for the $\kappa^{++} \kappa^{--}$ channel. In this subsection, we describe in detail the technical setup, including the Monte Carlo event generation and the detector simulation setup. For the background processes, the main contribution arises from the production of two massive gauge bosons. The $2$ muons plus missing energy signature receives background contributions from the production of $W^+ W^-$, $W^\pm Z$,  and $ZZ$. For all these backgrounds, we consider both the production through the $\mu^+ \mu^-$ annihilation and the VBF fusion, which is estimated using the prescription of Ref. \cite{Costantini:2020stv}. The cross sections for the $WW/WZ$ and $ZZ$ processes, including the decays into muons and neutrinos, are $0.65$ fb and $0.06$ fb,  respectively. In the case of four muon signature relevant to the pair production of doubly-charged scalars, the only background contribution arises from the production of $ZZ$, which can be categorized into two categories: $s$-channel like production of $ZZ$ and VBF production of $ZZ$ with cross sections of order $10^{-3}$--$10^{-1}$ fb including their decay branching ratios into muons\footnote{The contribution from the production of three and four gauge boson processes is subleading. On the other hand, the contribution of the backgrounds from the production of $\tau$ leptons decaying into $\mu\nu_\tau \nu_\mu$ is much smaller given the isolation criteria that we have defined to select muon candidates. Therefore, these subleading contributions will not be considered in this study.}. Events are generated using \textsc{Madgraph\_aMC@NLO} version 3.4.2 \cite{Alwall:2014hca} and then passed to \textsc{Pythia}~version 8310 \cite{Bierlich:2022pfr} to add parton showering and hadronization. To reduce the effects of statistical fluctuations in the tails of the kinematical distributions, we have generated about $3$--$5$ million events for the background processes depending on the respective cross sections. To take into account detector effects, we use the Simplified Fast Simulator (SFS) module \cite{Araz:2020lnp} in \textsc{MadAnalysis}~5 \cite{Conte:2012fm,Dumont:2014tja,Conte:2014zja,Conte:2018vmg,Araz:2019otb,Araz:2021akd}. The smearing and identification efficiencies of charged leptons and photons are implemented using the detector projection shipped with \textsc{Delphes} version 3.4.0 \cite{deFavereau:2013fsa}\footnote{More details about the identification and resolution maps for this study can be found in Ref. \cite{Belfkir:2023vpo}.}. For charged muons, we require them to be tightly isolated in the sense that the sum of the transverse momenta of all the tracks within $\Delta R = 0.2$ of the muon candidate excluding the muon itself to satisfy:
\begin{eqnarray}
    I_\mu \equiv \sum_{i \in {\rm tracks}} p_T^i < 0.1 \times p_T^\mu.
\end{eqnarray}
To estimate the sensitivity reach, we compute the signal significance defined using the Asimov formula \cite{Cowan:2010js}
\begin{eqnarray}
    {\cal S} \equiv \sqrt{2 \bigg((n_s + n_b) \log\bigg(1 + \frac{n_s}{n_b}\bigg) - n_s\bigg)}.
    \label{eq:SS:1}
\end{eqnarray}
On the other hand, we also consider the case where the background yields have some uncertainty denoted by $\delta$. In that case, the significance formula in equation \ref{eq:SS:1} becomes
\begin{eqnarray}
{\cal S}^{\delta[\%]} &=&\sqrt{2}\left[(n_s+n_b)\log\left(\frac{(n_s+n_b)(n_b+\delta_b^2)}{n_b^2+(n_s+n_b)\delta_b^2}\right) - \frac{n_b^2}{\delta_b^2} \log\left(1+\frac{\delta_b^2 n_s}{n_b(n_b+\delta_b^2)}\right)\right]^{1/2},
\label{eq:SS:2}
\end{eqnarray}
where $n_s$ and $n_b$ refer to the number of signal and background events, respectively, and $\delta_b = x n_b$ is the uncertainty on the background yield.

\subsection{Sensitivity reach in the $\phi^+\phi^-$ channel}

\subsubsection{Parton level}

We start by the pair production of the singly-charged scalars at future muon colliders. From the choices of the parameters in the different benchmark points, we found that the outcome of BP1, BP3 and BP5 is the same since they correspond to $m_\phi = 1.25$ TeV. The corresponding branching ratios of the charged scalars are quite similar in these three benchmark points (see table \ref{tab:BSs:BRs}). Therefore, we will study the sensitivity reach for BP1, BP2, and BP4 at $\sqrt{s} = 10$ TeV. At the parton level, we consider three final states: $\mu^\pm e^\mp + E_{T}^{\rm miss}$, $\mu^\pm \tau^\mp + E_{T}^{\rm miss}$ and $\mu^\pm \mu^\mp + E_{T}^{\rm miss}$. The dominant backgrounds to these production channels are the diboson production: $W^+ W^-$ and $Z Z$ where in the first background, both the two gauge bosons decay leptonically while in the second, only one of them decays leptonically and the other decays invisibly. The number of signal and background events is defined for the $\mu^\pm e^\mp + E_{T}^{\rm miss}$ as 
\begin{eqnarray}
N_s &=& 2 \times {\cal L} \times \sigma(\phi^+ \phi^-) \times \bigg[{\rm BR}(\phi^+ \to \mu^+ \bar{\nu}_\tau)~{\rm BR}(\phi^- \to e^- \nu_\mu) + {\rm BR}(\phi^+ \to \mu^+ \bar{\nu}_e)~{\rm BR}(\phi^- \to e^- \nu_\mu)\bigg], \nonumber \\
N_b &=& 2 \times {\cal L} \times \sigma(W^+ W^-) \times \bigg[{\rm BR}(W^+ \to \mu^+ \bar{\nu}_\mu)~{\rm BR}(W^- \to e^- \nu_e)\bigg],
\end{eqnarray}
where the factor $2$ is included to take into account combinatorics. Similarly, we can define the number of signal and background events for the case of the $\mu^\pm \tau^\mp + E_{T}^{\rm miss}$ channel as 
\begin{eqnarray}
N_s &=& 2 \times {\cal L} \times \sigma(\phi^+ \phi^-) \times \bigg[{\rm BR}(\phi^+ \to \mu^+ \bar{\nu}_\tau)~{\rm BR}(\phi^- \to \tau^- \nu_\mu) + {\rm BR}(\phi^+ \to \mu^+ \bar{\nu}_e)~{\rm BR}(\phi^- \to \tau^- \nu_\mu)\bigg], \nonumber \\
N_b &=& 2 \times {\cal L} \times \sigma(W^+ W^-) \times \bigg[{\rm BR}(W^+ \to \mu^+ \bar{\nu}_\mu)~{\rm BR}(W^- \to \tau^- \nu_e)\bigg],
\end{eqnarray}
and of the $\mu^\pm \mu^\mp + E_{T}^{\rm miss}$ channel as 
\begin{eqnarray}
N_s &=& {\cal L} \times \sigma(\phi^+ \phi^-) \times \bigg[{\rm BR}(\phi^+ \to \mu^+ \bar{\nu}_\tau)^2 + {\rm BR}(\phi^+ \to \mu^+ \bar{\nu}_e)^2 + 2 \times {\rm BR}(\phi^+ \to \mu^+ \bar{\nu}_\tau)~{\rm BR}(\phi^- \to \mu^- \nu_e)\bigg], \nonumber \\
N_b &=& {\cal L} \times \bigg[\sigma(W^+ W^-) {\rm BR}(W^+ \to \mu^+ \bar{\nu}_\mu)^2 + 2\times \sigma(ZZ) \times {\rm BR}(Z \to \mu^+ \mu^-)~{\rm BR}(Z \to \nu_\ell \bar{\nu}_\ell)\bigg],
\end{eqnarray}
where in the calculation of the number of events for the signal case we have not included decay channels with branching ratios smaller than $4\%$. \\

\begin{figure}[!t]
    \centering
    \includegraphics[width=0.325\linewidth]{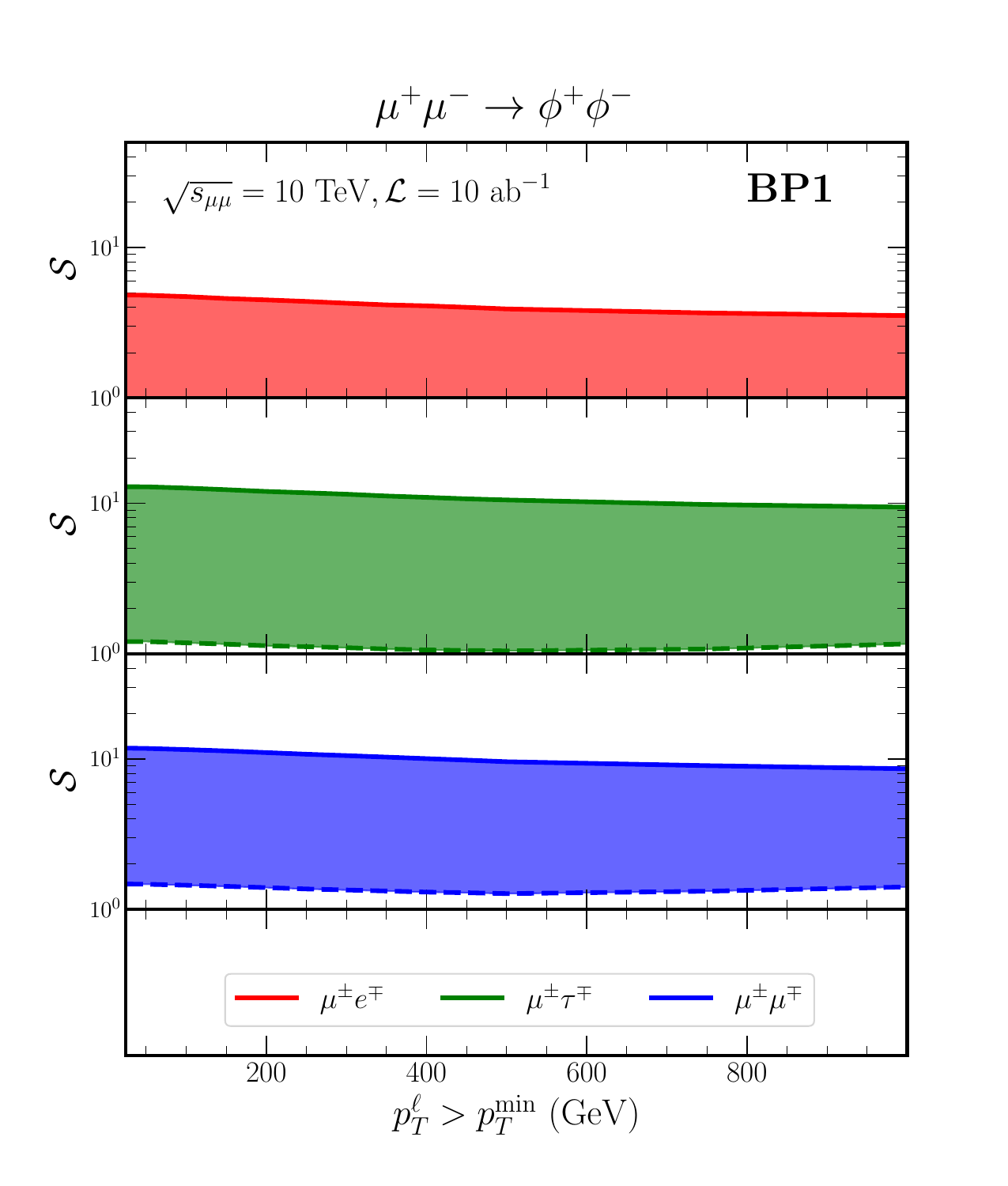}
    \hfill
    \includegraphics[width=0.325\linewidth]{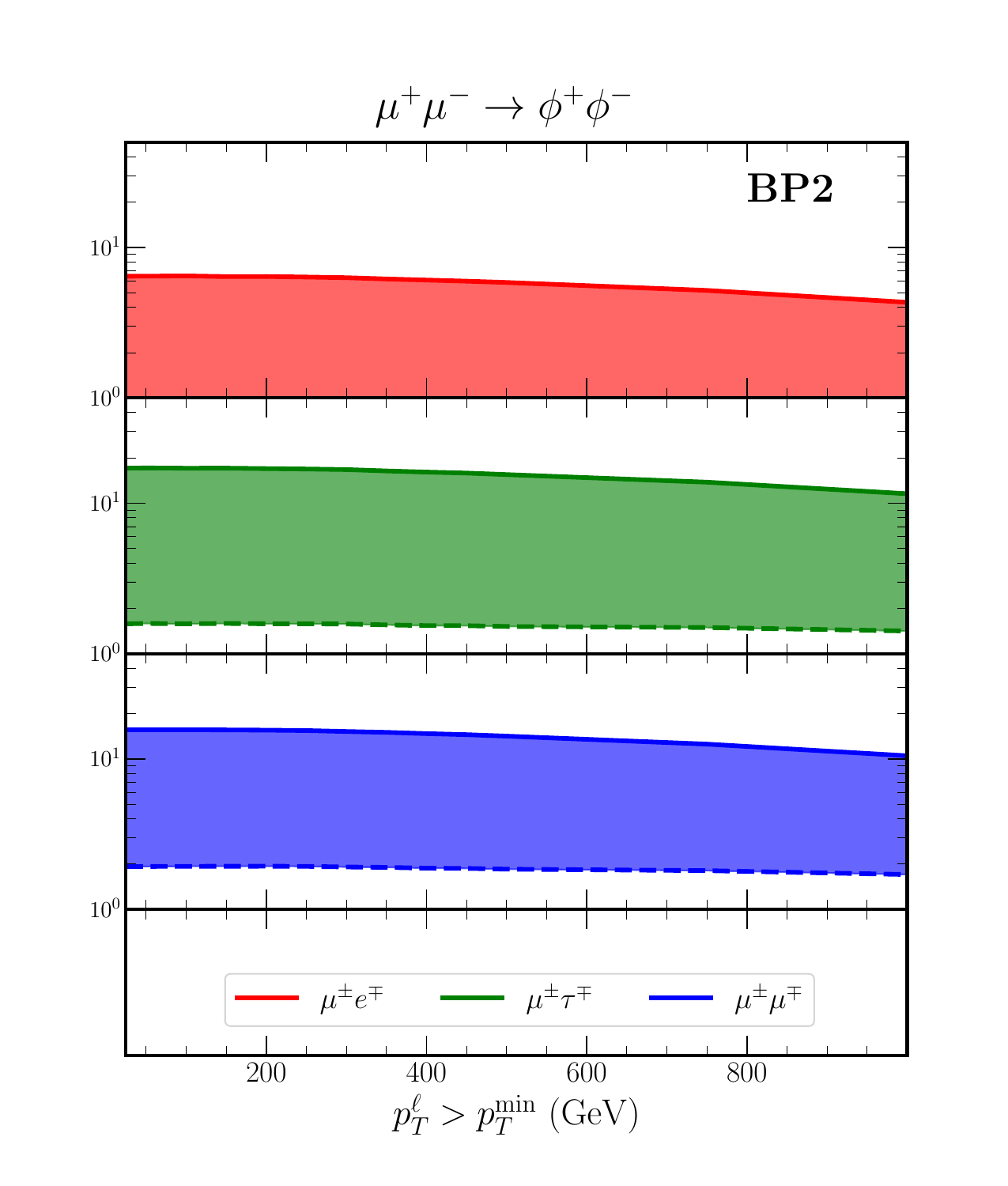}
    \hfill
    \includegraphics[width=0.325\linewidth]{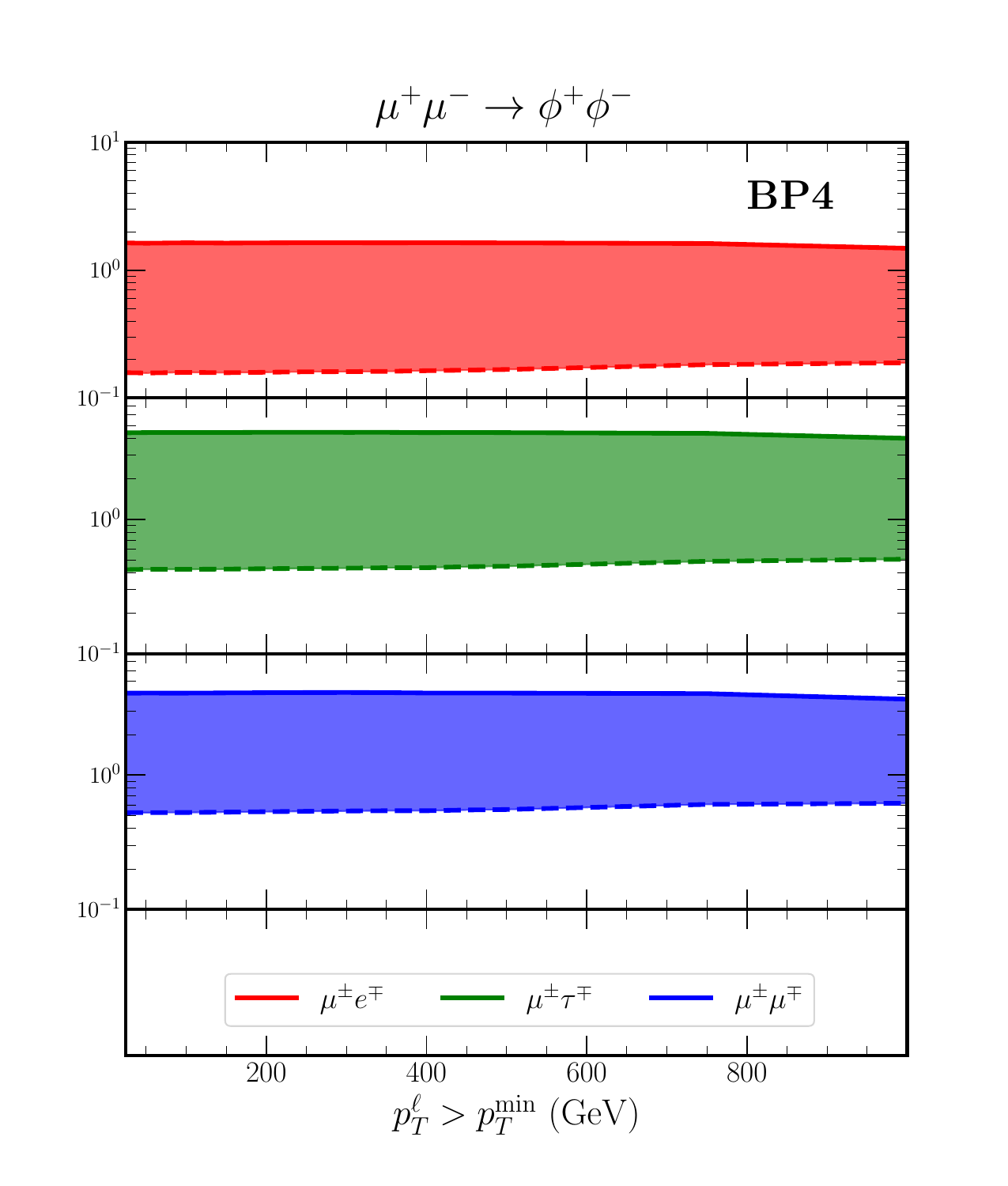}
    \caption{The signal significance as a function of the cut on the transverse momentum of the charged leptons for BP1 (left panel), BP2 (middle panel) and BP4 (right panel). Here we show the sensitivity for the $\mu^\pm e^\mp + E_{T}^{\rm miss}$ (red), $\mu^\pm \tau^\mp + E_{T}^{\rm miss}$ (green), and $\mu^\pm \mu^\mp + E_{T}^{\rm miss}$ (blue). The solid and dashed lines for each BP and each channel represents the result without and with $20\%$ uncertainty on the background yields.}
    \label{fig:significance:ss:parton}
\end{figure}

For this analysis, we have required the leptons to have at least $p_T > 25$ GeV and $|\eta| < 2.5$, which removes the contribution of charged leptons produced in the forward and backward regions. We also required that the magnitude of the missing transverse energy to be at least $50$ GeV. We then scanned over the threshold of selection on the transverse momentum of the charged lepton from $25$ GeV to $1000$ GeV and computed the significance for each selection threshold, including a $5\%$ uncertainty on the background events. The results are shown in figure \ref{fig:significance:ss:parton}. We can see that the signal significance can easily reach $5$ for the $\mu^\pm \tau^\mp$ and $\mu^\pm \mu^\mp$ channels at $10$ TeV. The sensitivity reach for the $\mu^\pm e^\mp$ channel is not so promising except in BP2.

\begin{figure}[!t]
    \centering
    \includegraphics[width=0.325\linewidth]{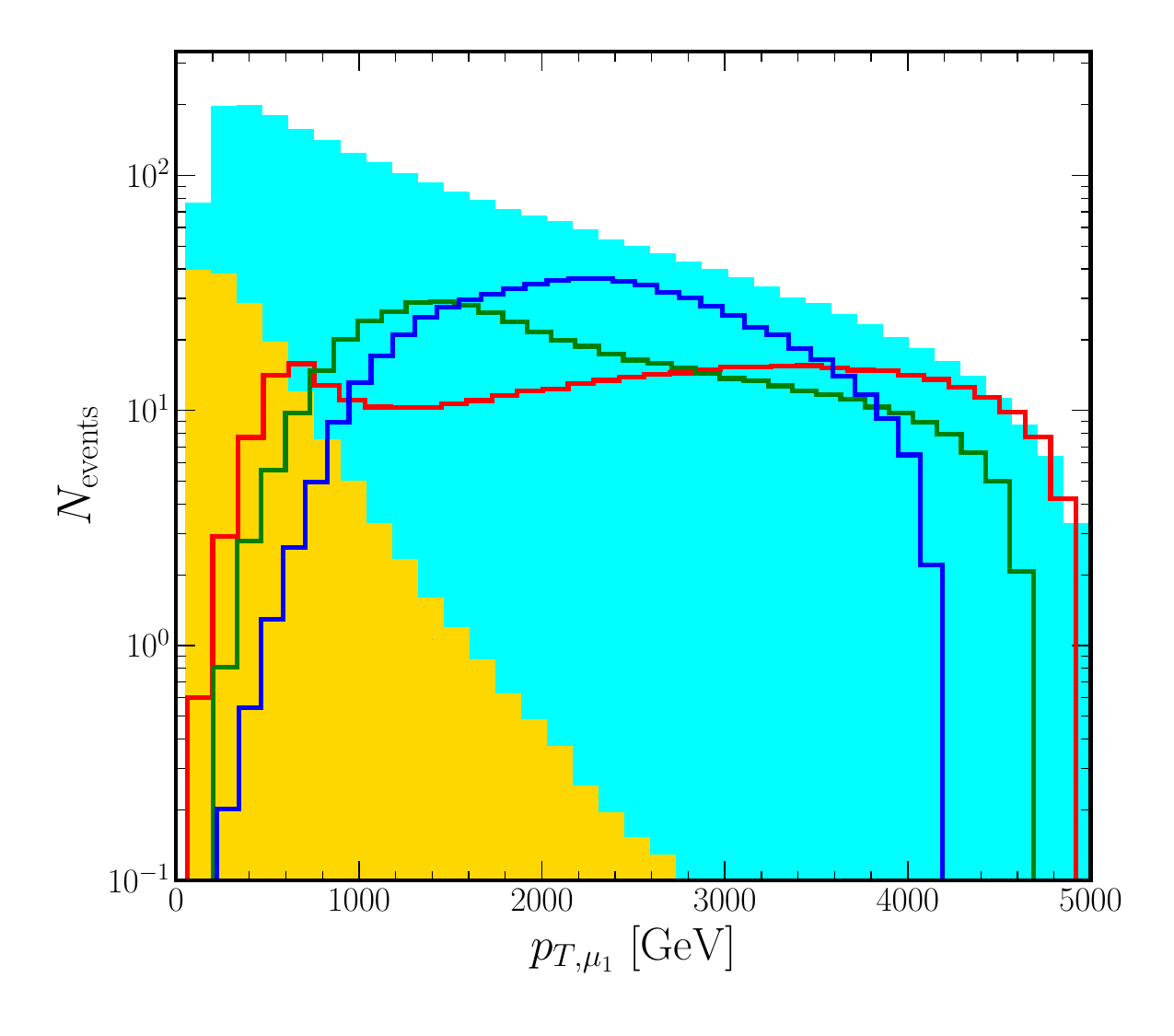}
    \hfill 
    \includegraphics[width=0.325\linewidth]{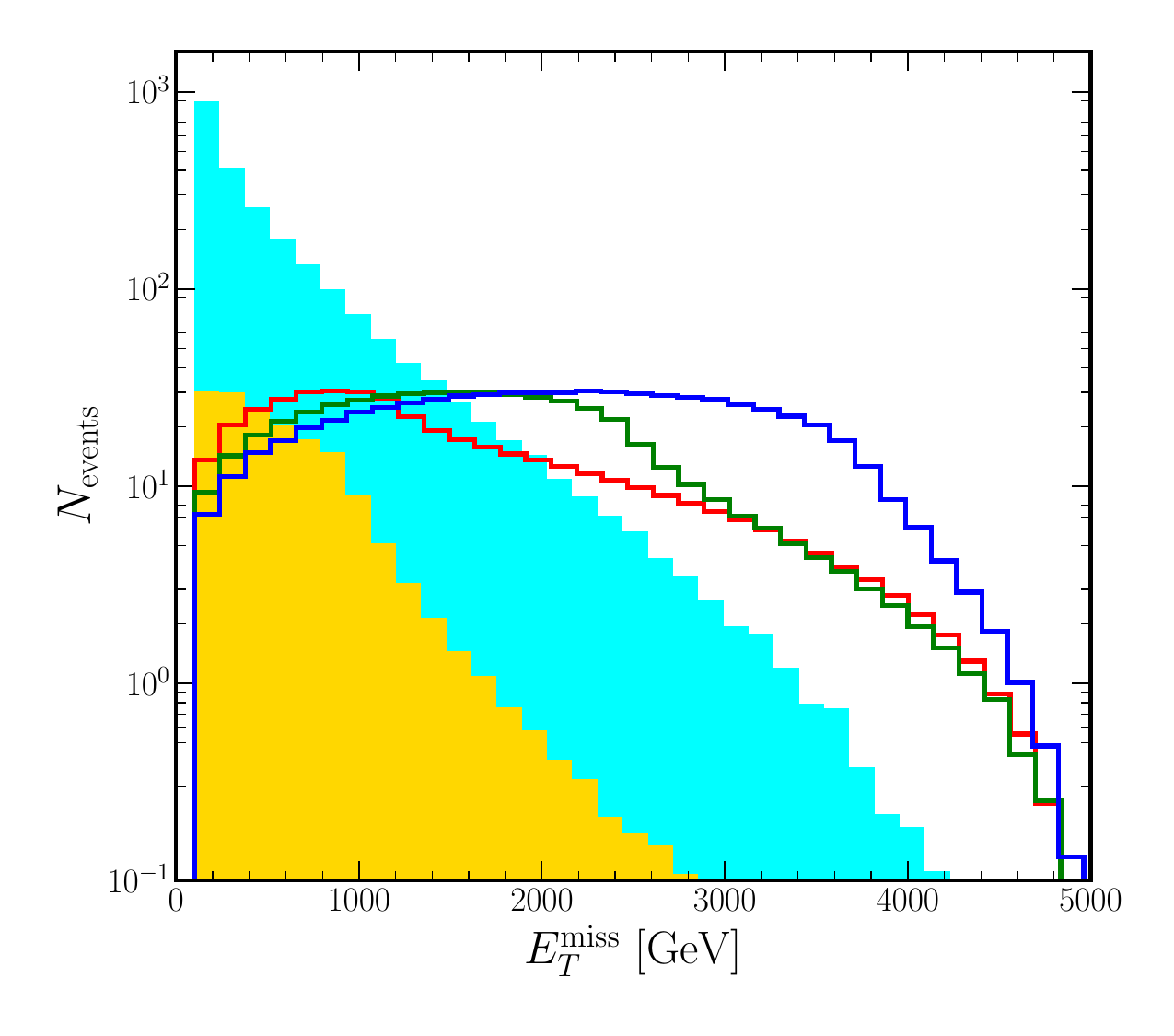}
    \hfill 
    \includegraphics[width=0.325\linewidth]{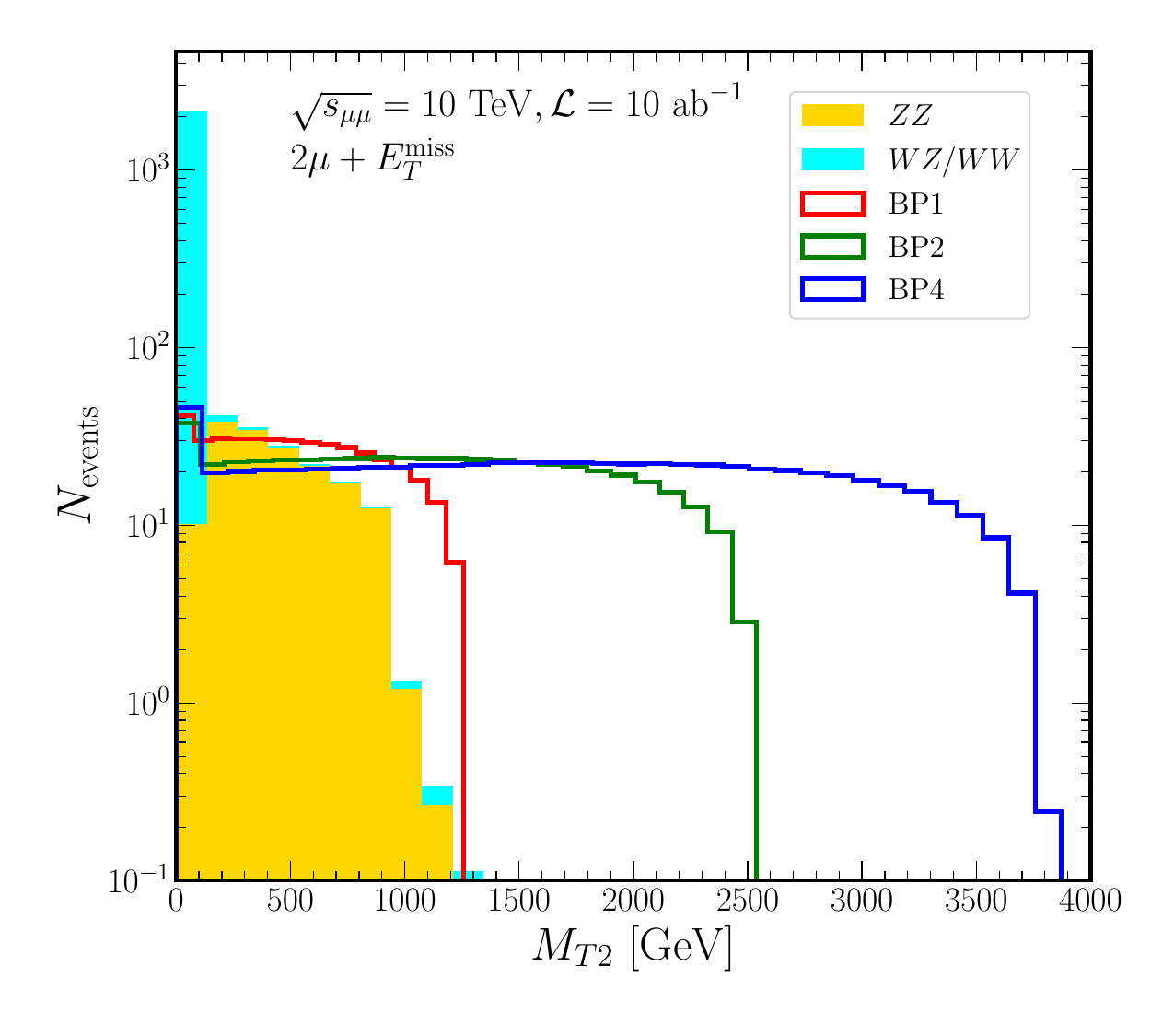}
    \vfill 
    \includegraphics[width=0.325\linewidth]{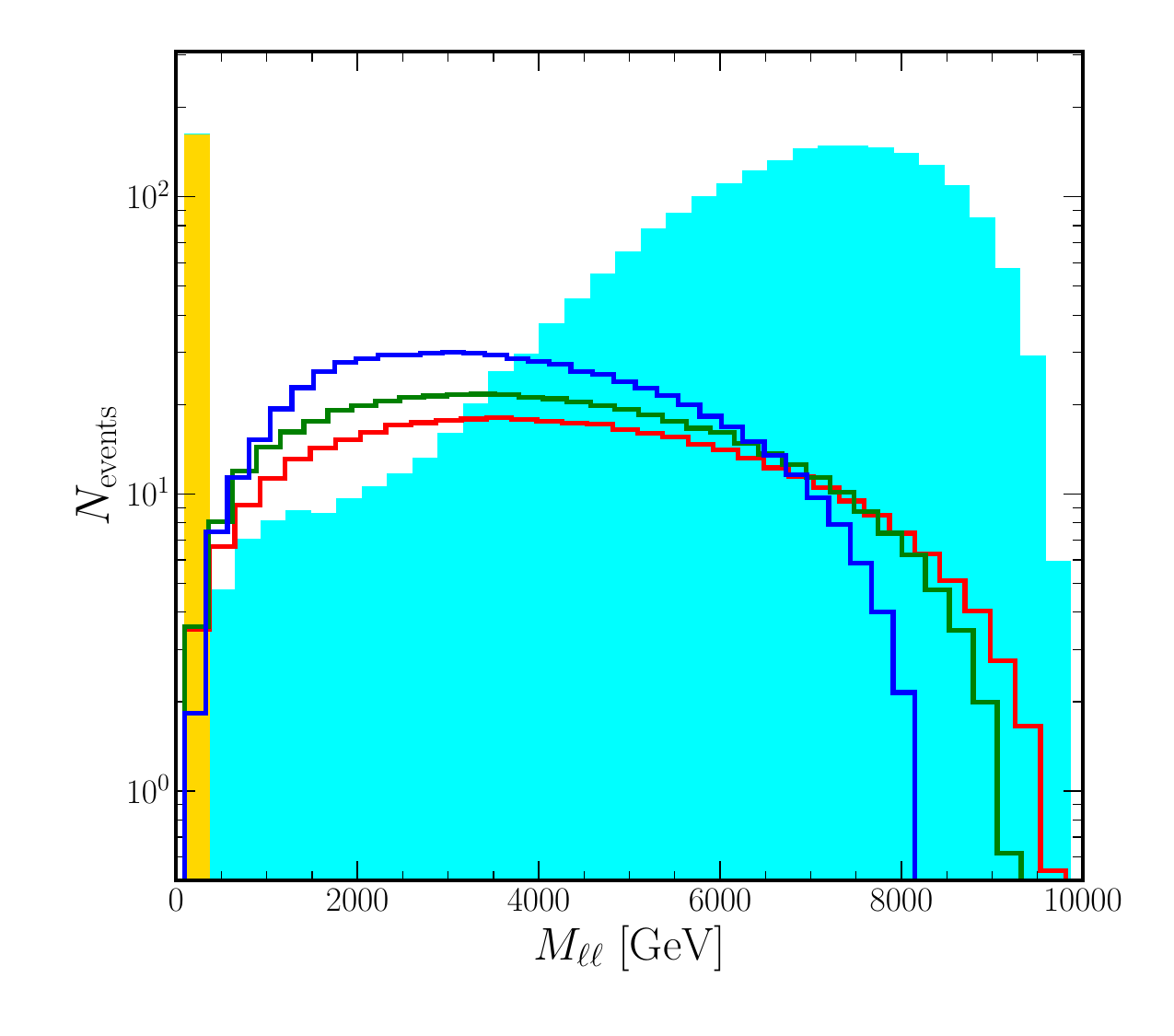} 
    \includegraphics[width=0.325\linewidth]{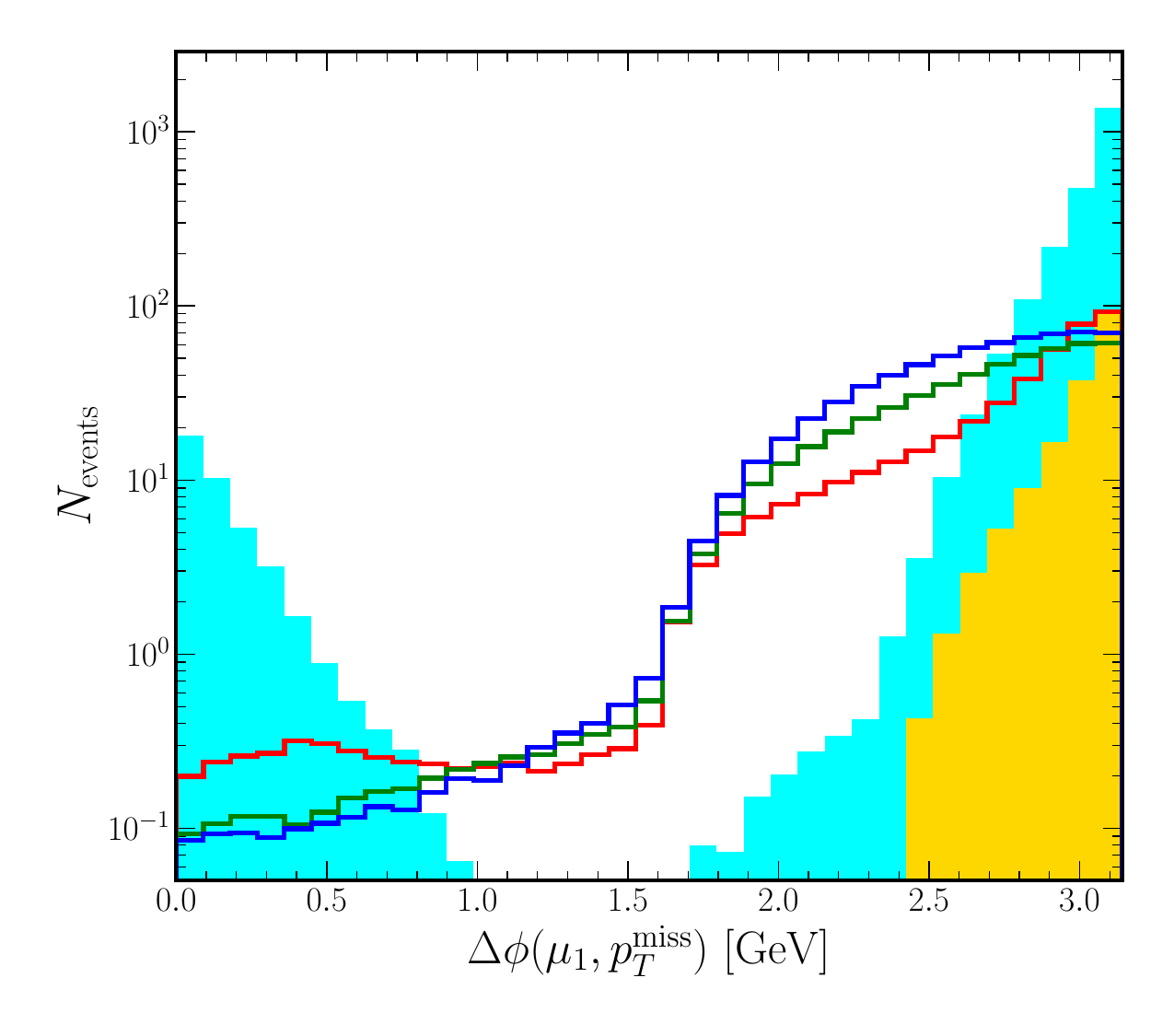}    
    \caption{Kinematic distributions for the two muons plus missing energy final state at $\sqrt{s}=10$ TeV and ${\cal L} = 10$ ab$^{-1}$. From top left to bottom right we show the transverse momentum of the leading muon ($p_{T,\mu_1}$), the transverse missing energy ($E_{T}^{\rm miss}$), the stransverse mass ($M_{T2}$), the invariant mass of dimuon system ($M_{\ell\ell}$) and the difference in the azimuthal angle between the leading muon and the missing energy momentum ($\Delta\phi(\mu_1, p_{\rm miss})$). The backgrounds are stacked on the top of each other where we show $ZZ$ (yellow) and $WW/WZ$ (cyan). The signal is shown for BP1 (red), BP2 (green), and BP4 (blue).}
    \label{fig:distributions:SS}
\end{figure}

\subsubsection{Detector level}
We now turn into a realistic analysis of the $\phi^+ \phi^-$ channel at the detector level in the 2 muons plus missing energy final state. The basic kinematic observables are shown in figure  \ref{fig:distributions:SS} while the cutflow table for the event selection is displayed in Table \ref{cutflow:SS}. First, we require events to satisfy $E_{T}^{\rm miss} > 100$ GeV. We also veto events that contain one isolated electron with $p_T > 10$ GeV and $|\eta| < 2.5$, one tau lepton with $p_T > 25$ GeV and $|\eta| < 2.5$ and one photon with $p_T > 100$ GeV and $|\eta| < 2.5$. Since the two singly-charged scalars decay exclusively into muons, we require the presence of exactly two muons with $p_T > 30$ GeV and $|\eta| < 2.5$. The two muons are required to have opposite electric charge. The $ZZ$ backgrounds can be further suppressed by requiring that the invariant mass of the two muon system to be larger than $100$ GeV. After this selection, about $50\%$ of $ZZ$ background events are removed while the contribution of $WW/WZ$ backgrounds remains unchanged. The stransverse mass ($M_{T2}$) is a useful kinematic variable that can be employed to reduce the contributions of the $WW$ backgrounds \cite{Lester:1999tx,Barr:2003rg,Lester:2011nj}. Denoting the two charged lepton momenta by $p_{T}^{(1)}$ and $p_{T}^{(2)}$, and the missing particle momenta as $q_{T}^{(1)}$ and $q_{T}^{(2)}$, the $M_{T2}$ variable is calculated as follows:
\begin{eqnarray}
M_{T2} = \min_{q_{T}^{(1)} + q_{T}^{(2)} = {p}_{\mathrm{miss}}}\bigg\{\max\{M_T(p_{T}^{(1)}, q_{T}^{(1)}), M_T(p_{T}^{(2)}, q_{T}^{(2)})\} \bigg\},
\end{eqnarray}
where $q_{T}^{(1)}$ and $q_{T}^{(2)}$ are combined to form the total missing momentum. The calculation of the $M_{T2}$ also relies on the test mass for the invisible particle, which we choose to be equal to zero. For this choice, the $M_{T2}$ distribution will have an end point at around the mass of the parent particle, which can be seen clearly in the top right panel of figure  \ref{fig:distributions:SS} where the end of point of $M_{T2}$ is at around $1.25$ TeV (BP1), $2.5$ TeV (BP2), and $3.75$ TeV (BP4). There are exceptions for the SM backgrounds since we have contributions of VBF diagrams that affect the expected behavior in $s$--channel mediated processes. We further require $M_{T2} > 100$ GeV, which is enough to suppress the $WW/WZ$ and $ZZ$ backgrounds by $99\%$ and $2\%$, respectively. To further reduce the contribution of $ZZ$ backgrounds, we also impose the condition $\Delta\phi(\vec{\mu}_1, \vec{p}_{\rm miss}) > 0.5$ where $\vec{\mu}_1$ is the 3-momentum of the leading muon. This selection reduces the yields of the two backgrounds by about $50\%$. Finally, we impose that $M_{\ell \ell}$ falls in the range $]200,~5000]$ GeV, which dramatically reduces the number of events for the $ZZ$ backgrounds. The final selection efficiency for the signal events is of order $44\%$--$56\%$ (higher for higher scalar masses). 

\begin{table}[!t]
\setlength\tabcolsep{5pt}
\begin{center}
\begin{adjustbox}{max width=1.01\textwidth}
    \begin{tabular}{l cc cc cc cc}
    \toprule
    \toprule
      & \multicolumn{2}{c}{$WW/WZ$} & \multicolumn{2}{c}{$ZZ$} & \multicolumn{2}{c}{${\rm BP1}$} & \multicolumn{2}{c}{${\rm BP2}$} \\ 
      \toprule
      \toprule
      & Events & $\varepsilon$ & Events & $\varepsilon$ & Events & $\varepsilon$ & Events & $\varepsilon$ \\ 
      \toprule
      Initial                                 & 6548.8 & -  & 641.0 & - & 482.5 & -  & 575.0 & -  \\
      $E_{T}^{\rm miss} > 100~{\rm GeV}$      & 1851.4 $ \pm $ 1.3 & 0.283 & 554.0 $ \pm $ 0.1 & 0.864 & 477.9 $ \pm $ 0.1 & 0.990 & 572.0 $ \pm $ 0.0 & 0.995 \\
      Electron Veto                           & 1842.8 $ \pm $ 1.3 & 0.995 & 470.0 $ \pm $ 0.2 & 0.848 & 475.7 $ \pm $ 0.1 & 0.995 & 569.5 $ \pm $ 0.1 & 0.996 \\
      $\tau$ Veto                             & 1842.5 $ \pm $ 1.3 & 1.000 & 470.0 $ \pm $ 0.2 & 1.000 & 475.7 $ \pm $ 0.1 & 1.000 & 569.4 $ \pm $ 0.1 & 1.000 \\
      Photon Veto                             & 1607.7 $ \pm $ 1.1 & 0.873 & 461.4 $ \pm $ 0.2 & 0.982 & 450.1 $ \pm $ 0.1 & 0.946 & 540.6 $ \pm $ 0.1 & 0.949  \\
      $n_\mu = 2$                             & 1327.5 $ \pm $ 1.0 & 0.886 & 213.6 $ \pm $ 0.1 & 0.474 & 417.0 $ \pm $ 0.2 & 0.964 & 505.0 $ \pm $ 0.2 & 0.972 \\
      2 SFOS                                  & 1327.5 $ \pm $ 1.0 & 1.000 & 213.3 $ \pm $ 0.1 & 0.999 & 417.0 $ \pm $ 0.2 & 1.000 & 505.0 $ \pm $ 0.2 & 1.000 \\ 
      $M_{\ell\ell} > M_Z$                    & 1327.5 $ \pm $ 1.0 & 1.000 & 98.8 $ \pm $ 0.1 & 0.463 & 416.8 $ \pm $ 0.2 & 1.000 & 505.0 $ \pm $ 0.2 & 1.000 \\
      $M_{T2} > 100~{\rm GeV}$                & 5.7 $ \pm $ 0.0 & 0.004 & 97.2 $ \pm $ 0.1 & 0.984 & 367.6 $ \pm $ 0.2 & 0.882 & 468.8 $ \pm $ 0.2 & 0.928 \\
      $\Delta\phi(\vec{\mu}_1, \vec{p}_{\rm miss}) > 0.5$& 3.3 $ \pm $ 0.0 & 0.999 & 58.7 $ \pm $ 0.0 & 1.000 & 340.8 $ \pm $ 0.2 & 0.998 & 449.4 $ \pm $ 0.2 & 0.999 \\
      $M_{\ell \ell} \in ~]200, 5000[~{\rm GeV}$& 1.2 $ \pm $ 0.0 & 0.362 & 0.2 $ \pm $ 0.0 & 0.004 & 214.5 $ \pm $ 0.2 & 0.630 & 298.0 $ \pm $ 0.2 & 0.663  \\
      \toprule
      \toprule
    \end{tabular}
    \end{adjustbox}
  \end{center}
      \caption{Cutflow tables for the 2 muons plus missing energy analysis. Here we show the two backgrounds ($WW/WZ$ and $ZZ$) and two benchmark points (BP1 and BP2). For each selection step we also calculate the uncertainty on the number of events and the efficiency defined as $\varepsilon = N_{i}/N_{i-1}$ where $N_k$ refers to the number of events after the selection step $k$. All the events are normalized to $10$ ab$^{-1}$ of Luminosity.}
      \label{cutflow:SS}
\end{table}

\begin{figure}[!t]
    \centering
    \includegraphics[width=0.5\linewidth]{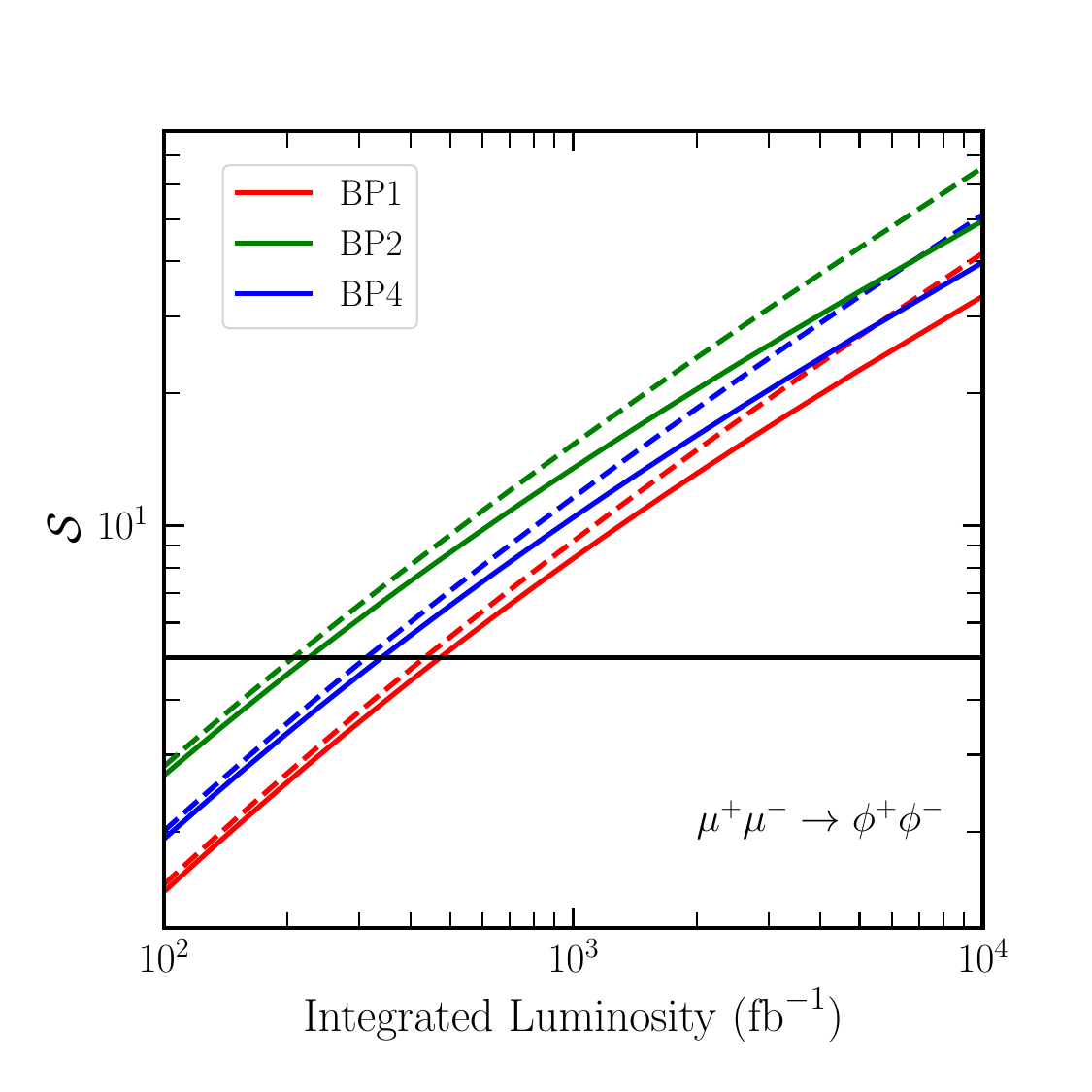}
    \caption{Signal significance for the discovery reach of the singly-charged scalar in the $\mu^+\mu^- + E_{T}^{\rm miss}$ final state as a function of the luminosity. Here we show the results for BP1 (red), BP2 (blue),  and BP4 (green). The results shown assuming $0\%$ and $20\%$ uncertainty on the background yields in dashed and solid lines, respectively.}
    \label{fig:SS:significance}
\end{figure}

We finally calculate the signal significance in the signal region defined as the last step of the selection in Table \ref{cutflow:SS}. In this  we assume that the number of backgrounds is the same for all the luminosities ($n_b\equiv N_{WW} + N_{ZZ} = 1.4$), and we take into account two assumptions on the background uncertainty, {\it i.e.} assuming that $\delta = 0\%$ (eq. \ref{eq:SS:1}) and $\delta = 20\%$ (eq. \ref{eq:SS:2}). The results are shown in figure  \ref{fig:SS:significance} where we can see that luminosities of about $200$--$400$ fb$^{-1}$ are enough to discover or exclude our signal benchmark points, therefore competing with or even outperforming the sensitivity reach of the HL-LHC. 

\subsection{Sensitivity reach in the $\kappa^{++}\kappa^{--}$ channel}

\subsubsection{Parton level}

The case of the pair production of doubly-charged scalars is much more simpler since they decay predominantly into $\mu\mu$ with branching ratios of $93.6\%$--$99.7\%$ except in the case of BP5 where the decay branching ratio into muons is $63.1\%$. Using similar arguments regarding mass choices in our benchmark points, we find that BP1, BP2, and BP4 will lead to similar consequences since they correspond to $m_\kappa = 1.25$ TeV. Therefore, in this channel, we will analyse the sensitivity reach in the following benchmark points: BP1, BP3, and BP5. The number of signal and background events is defined as 
\begin{eqnarray}
N_s &=& {\cal L} \times \sigma(\kappa^{++} \kappa^{--}) \times {\rm BR}(\kappa^{++} \to \mu^+ \mu^+)^2, \nonumber \\
N_b &=& {\cal L} \times \sigma(ZZ) \times {\rm BR}(Z \to \mu^+ \mu^-)^2.
\end{eqnarray}

It is clear that we can easily reach signal significance larger than ${\cal O}(1)$ given the smallness of the associated background cross sections. We do not include the backgrounds from vector-boson fusion since they involve missing energy and extra leptons, and therefore, their inclusion would require a dedicated analysis strategy, which we will discuss in the next section. We first select events that contain exactly four muons with transverse momentum of at least $25$ GeV and pseudorapidity smaller than $2.5$. We also require that muons with opposite sign to be combined with each other to form a Z--boson {\it candidate}. 

\begin{figure}[!t]
    \centering
    \includegraphics[width=0.325\linewidth]{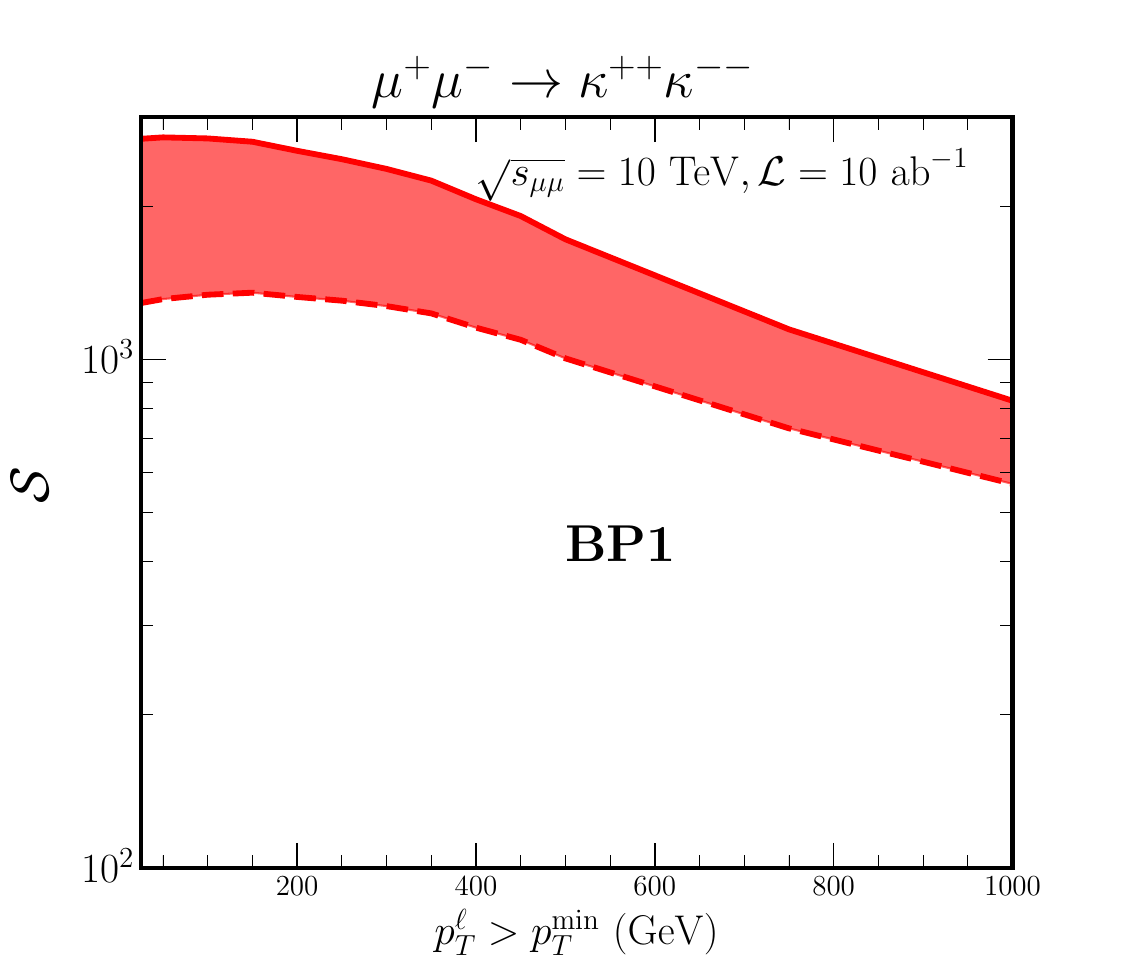}
    \hfill
    \includegraphics[width=0.325\linewidth]{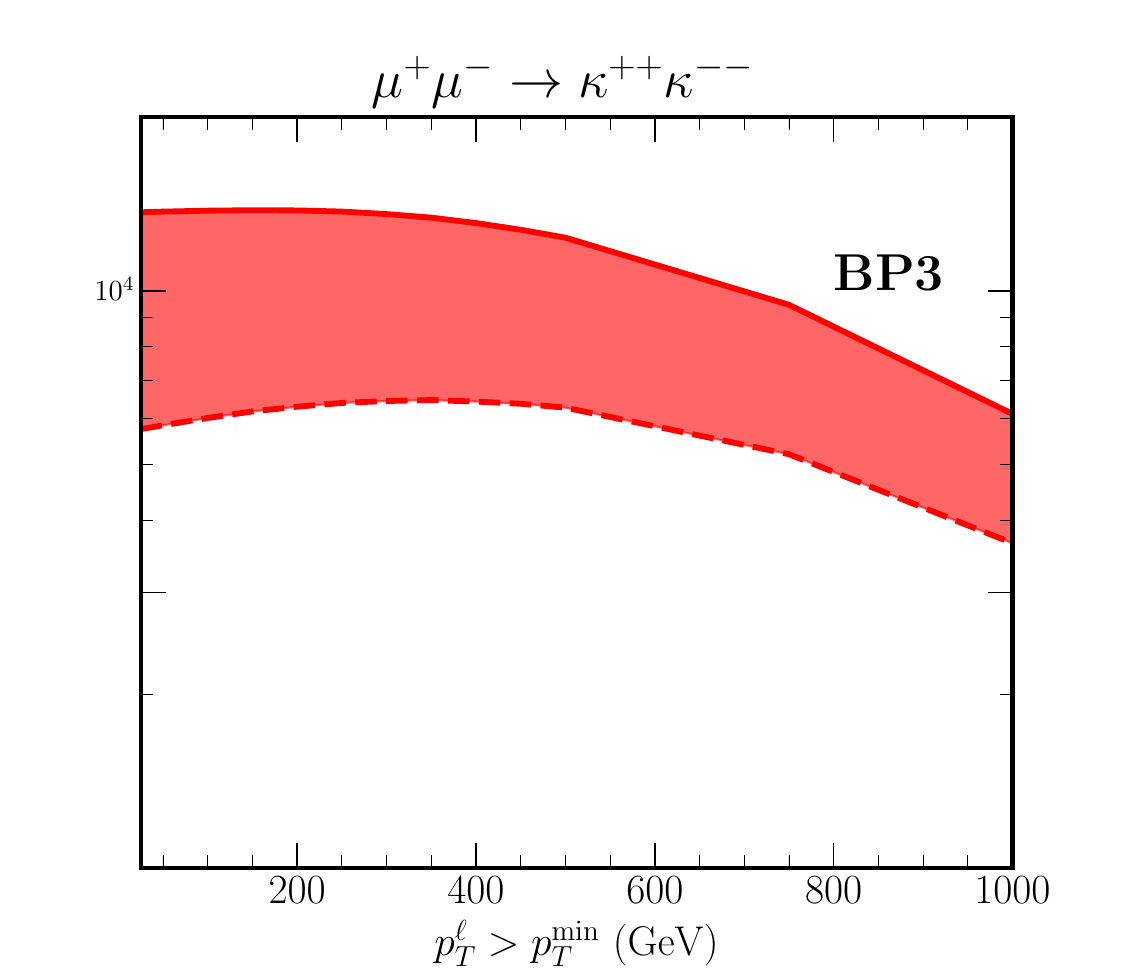}
    \hfill
    \includegraphics[width=0.325\linewidth]{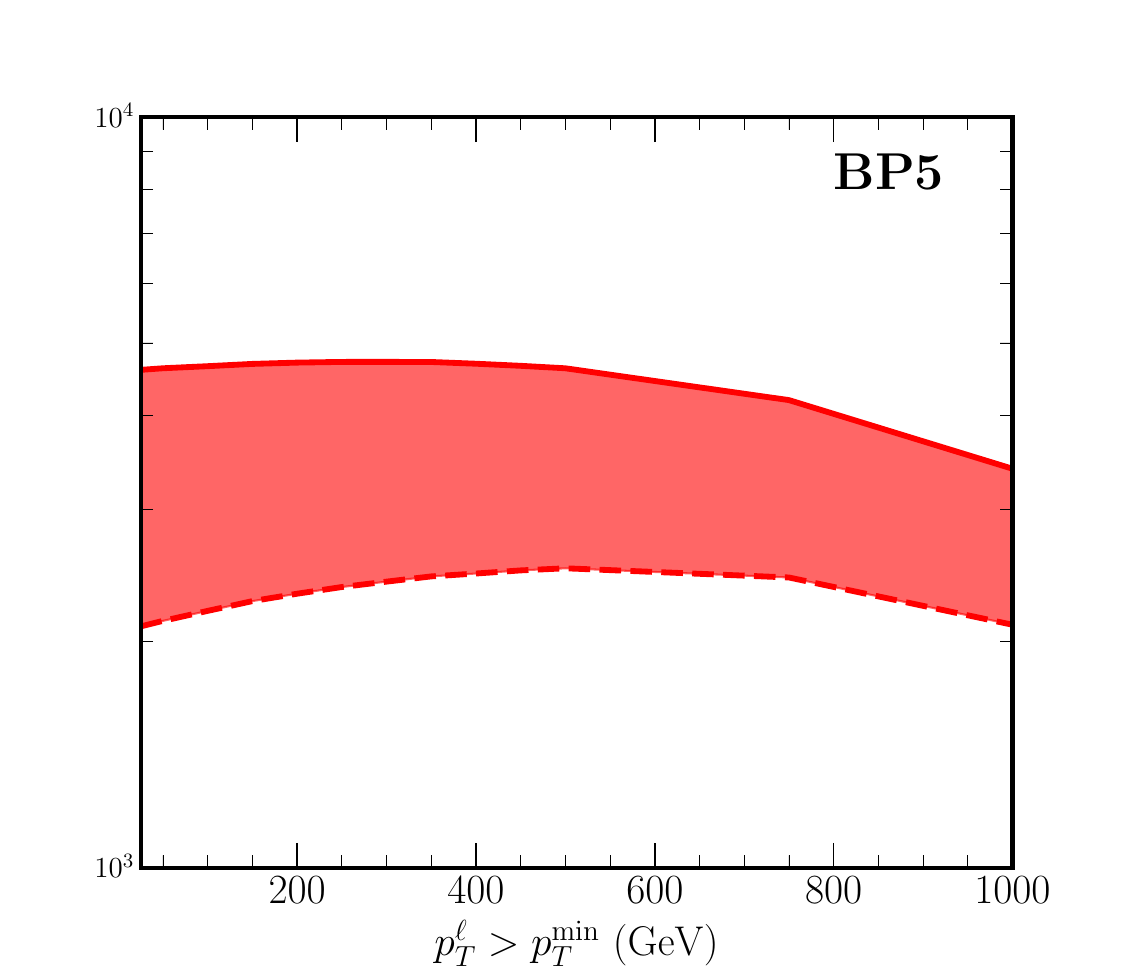}
    \caption{Signal significance as a function of the cut on the transverse momentum of the charged lepton for the $\kappa^{++} \kappa^{--}$ channel for BP1 (left panel), BP3 (middle panel),  and BP5 (right panel).}
    \label{fig:significance:kk:parton}
\end{figure}

To further reduce the background contribution, we require that the invariant mass of these $Z$--boson candidates to be larger than $100$ GeV (which kills most of the $Z$--boson contribution to the background). We then scan over the minimum transverse momentum of the muon for the range of $25$--$1000$ GeV and compute the signal significance as defined in equations \ref{eq:SS:1}--\ref{eq:SS:2} including a $20\%$ background uncertainty (dashed lines). The results are shown in figure \ref{fig:significance:kk:parton}, where we can see that the signal significance is very large for all the benchmark points. We conclude that the muon colliders will have a great potential to discover the Zee-Babu model, at least in channels involving doubly-charged scalars.  We notice the importance of a dedicated analysis involving state-of-art Monte Carlo tools to comprehensively study the model at muon colliders, especially concerning reconstructions of scalars at muon colliders and their characterization. This will be discussed in the next section.

\begin{table}[!t]
\setlength\tabcolsep{5pt}
\begin{center}
\begin{adjustbox}{max width=1.01\textwidth}
    \begin{tabular}{l cc cc cc cc}
    \toprule
    \toprule
      & \multicolumn{2}{c}{$ZZ~({\rm s}$-${\rm ch})$} & \multicolumn{2}{c}{$ZZ~({\rm VBF})$} & \multicolumn{2}{c}{${\rm BP1}$} & \multicolumn{2}{c}{${\rm BP3}$} \\
    \toprule
    \toprule
      & Events & $\varepsilon$ & Events & $\varepsilon$ & Events & $\varepsilon$ & Events & $\varepsilon$ \\ \hline
      Initial                                 & 37.4 & -  & 2352.9 & - & $6.4 \times 10^{5}$ & - & $7.5 \times 10^{6}$ & -  \\
      $E_{T}^{\rm miss} < 100~{\rm GeV}$      & 31.5 $ \pm $ 0.0 & 0.843 & 142.5 $ \pm $ 0.1 & 0.061 & $5.0 \times 10^{5}$ $ \pm $ 264.8 & 0.775 & $5.7 \times 10^{6}$ $ \pm $ 3137.0 & 0.759 \\
      Electron Veto                           & 31.5 $ \pm $ 0.0 & 0.999 & 142.4 $ \pm $ 0.1 & 0.999 & $5.0 \times 10^{5}$ $ \pm $ 266.2 & 0.993 & $5.7 \times 10^{6}$ $ \pm $ 3149.1 & 0.993 \\
      $\tau$ Veto                             & 31.5 $ \pm $ 0.0 & 1.000 & 142.4 $ \pm $ 0.1 & 1.000 & $5.0 \times 10^{5}$ $ \pm $ 266.2 & 1.000 & $5.7 \times 10^{6}$ $ \pm $ 3149.3 & 1.000 \\
      Photon Veto                             & 30.9 $ \pm $ 0.0 & 0.982 & 141.7 $ \pm $ 0.1 & 0.995 & $4.4 \times 10^{5}$ $ \pm $ 276.8 & 0.896 & $5.2 \times 10^{6}$ $ \pm $ 3233.0 & 0.919 \\
      Jet Veto                                & 29.8 $ \pm $ 0.0 & 0.964 & 137.0 $ \pm $ 0.1 & 0.967 & $3.9 \times 10^{5}$ $ \pm $ 273.2 & 0.868  & $4.7 \times 10^{6}$ $ \pm $ 3222.1 & 0.902 \\
      $n_\mu = 4$                             & 3.7 $ \pm $ 0.0 & 0.125 & 12.9 $ \pm $ 0.0 & 0.094 & $3.8 \times 10^{5}$ $ \pm $ 272.3 & 0.988 & $4.7 \times 10^{6}$ $ \pm $ 3218.1 & 0.993 \\
      2 SFSS                                  & 3.7 $ \pm $ 0.0 & 1.000 & 12.9 $ \pm $ 0.0 & 1.000 & $3.8 \times 10^{5}$ $ \pm $ 272.3 & 1.000 & $4.7 \times 10^{6}$ $ \pm $ 3218.1 & 1.000 \\
      $M_{\mu\mu} > 1000~{\rm GeV}$           & 3.6 $ \pm $ 0.0 & 0.971 & 1.3 $ \pm $ 0.0 & 0.104 & $3.6 \times 10^{5}$ $ \pm $ 268.1 & 0.951 & $4.7 \times 10^{6}$ $ \pm $ 3216.1 & 0.997 \\
      \toprule
      \toprule
    \end{tabular}
    \end{adjustbox}
  \end{center}
   \caption{Same as in Table \ref{cutflow:SS} but for the case of the four muon final state.}
   \label{cutflow:KK}
\end{table}

\begin{figure}[!t]
    \centering
    \includegraphics[width=0.49\linewidth]{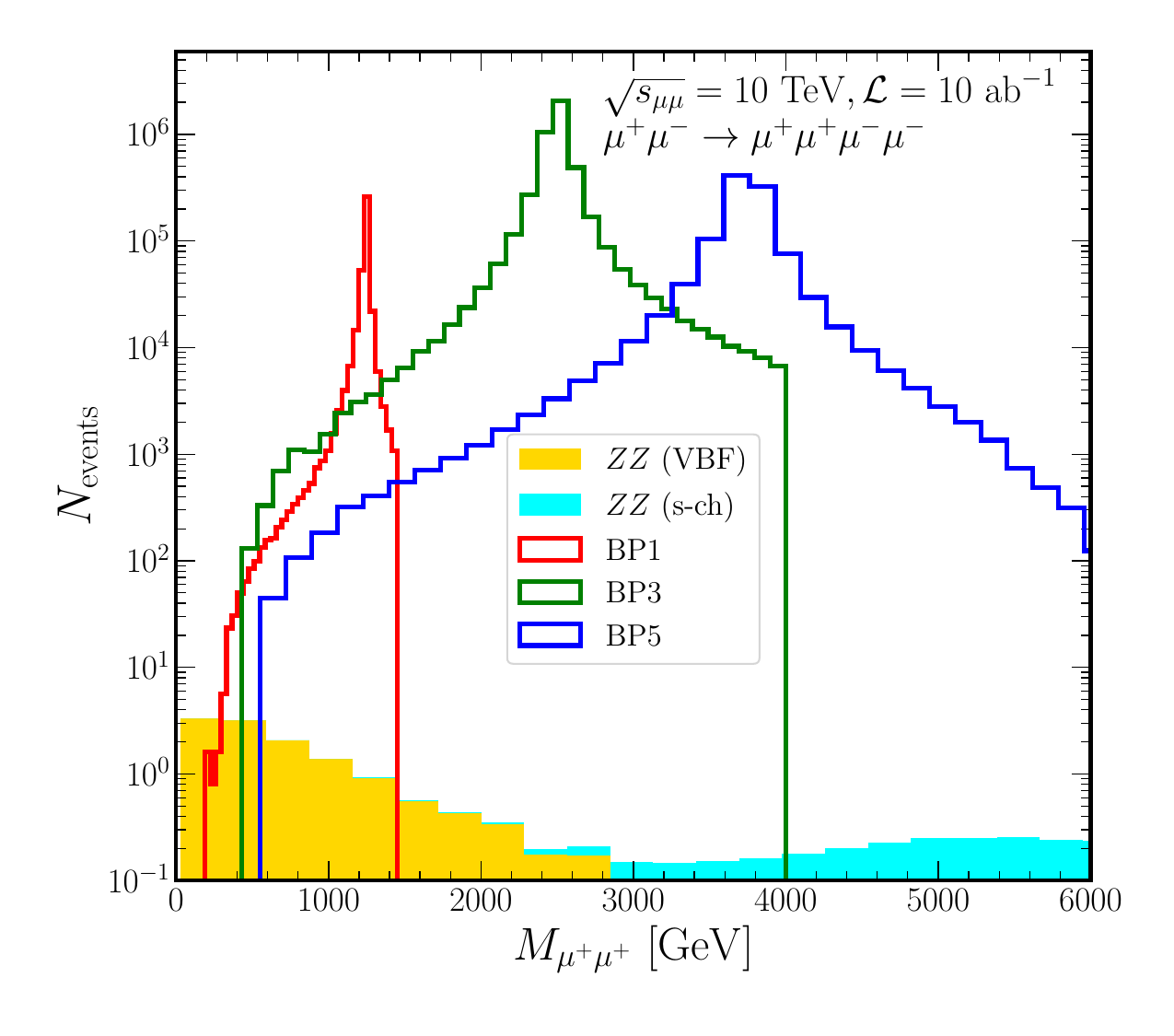}
    \caption{Invariant mass distribution of the $\mu^+ \mu^+$ for the backgrounds (yellow and cyan) and three benchmark points of the signal (BP1, BP3,  and BP5)}
    \label{fig:Mass:Kappa}
\end{figure}

\subsubsection{Detector level}

We now turn to the analysis at the detector level using the same technical setup discussed in section \ref{sec:technical}. The contribution of the $ZZ$ production through VBF can be drastically reduced by imposing cuts on missing energy (in case of charged-current production) and requiring exactly four muons (in case of neutral-current production). The cutflow table for this channel is shown in Table \ref{cutflow:KK}. We first select events if the total transverse missing energy is smaller than $100$ GeV. Such a choice is motivated by the fact that the signal process does not involve any missing energy besides some minor contribution from misidentification of muon candidates. This selection step reduces the contribution of VBF $ZZ$ by $93\%$ while slightly affecting both the $ZZ$ contribution through $s$--channel and the signal events for all the benchmark points. We then apply a number of vetoes on several objects like electrons with $p_T > 15$ GeV and $|\eta| < 2.5$, $\tau$ leptons with $p_T > 25$ GeV and $|\eta| < 2.5$, photons with $p_T > 100$ GeV and jets with $p_T > 30$ GeV and $|\eta| < 2.5$.  We then require the existence of four charged muons forming two pairs of same electric charge. Those pairs will be used to form doubly-charged scalar candidates. The invariant mass of these candidates is required to be larger than $1000$ GeV. The final efficiency for the backgrounds is small ($0.09$ for $s$--channel $ZZ$ and $5.5 \times 10^{-4}$ for VBF $ZZ$). On the other hand, the signal efficiency is $> 50\%$. To check the ability of our analysis in reconstructing the invariant mass of the $\kappa$ candidates, we display the invariant mass of the system formed by same sign same flavor (SSSF) muons in figure \ref{fig:Mass:Kappa} for BP1 (red), BP3 (blue) and BP5 (green). We can see from that the peak of the invariant mass distribution is centered around the mass of the doubly-charged scalar candidate, and the width of the distribution grows with the mass reflecting the full off-shell effects taken into account in our simulation. Given the large number of events that survive the final selection, we can conclude that the discovery reach for the case of $\kappa^{++} \kappa^{--}$ is even much more promising than the case of $\phi^+ \phi^-$ production.

\section{Summary and conclusions}\label{sec:conclusions}

In this work, we have studied the phenomenology of the Zee-Babu model at future muon colliders. This model provides a natural explanation of the smallness of neutrino mass through two-loop radiative neutrino mass generation mechanism and with the exchange of two new $SU(2)_L$ singlet scalars: a singly-charged and a doubly-charged scalars. After studying the impact of the constraints from neutrino oscillation data, lepton flavour violation, and lepton flavour universality tests, we have selected five phenomenologically viable benchmark points which depend on the masses of the charged scalars of the model. We then analyzed the production rates for various channels that include charged leptons (both FC and FV channels), singly-charged scalars and doubly-charged scalars for the five benchmark points. A special attention was given to channels that not only probe the Yukawa-type couplings connected to the lepton sector but also channels that are sensitive to the nature of the scalar potential, in particular, the lepton-number violating coupling. These channels involve $n \geq 4$ scalars and are enjoying almost a background-free environment even that the production cross sections are of order $10^{-3}$--$10^{0}$ fb. In particular, the production of four doubly-charged scalars lead to a golden channel consisting of eight highly-energetic muons. We finally analyzed the signal-to-background for the pair production of two charged scalars: $\phi^+ \phi^-$ and $\kappa^{++} \kappa^{--}$ at $\sqrt{s}=10$ TeV both at the parton level and the detector level.  For the pair production of singly-charged scalars ($\phi^+ \phi^-$), we have analyzed three channels, {\it i.e.} $e^\pm \mu^\mp + E_{T}^{\rm miss}$, $\mu^\pm \tau^\mp + E_{T}^{\rm miss}$ and $\mu^\pm \mu^\mp + E_{T}^{\rm miss}$. For the pair production of doubly-charged scalars, we have analyzed the production of $\mu^+\mu^-\mu^+\mu^-$ and we found even more promising sensitivity. 

\section*{Acknowledgements}
The work of A.J. is supported by the Institute for Basic Science (IBS) under the project code, IBS-R018-D1.  The  work of S.N is supported by the United Arab Emirates University (UAEU) under UPAR Grant No. 12S093. S.S. would like to thank Admir Greljo for discussion. 

\bibliographystyle{style}
\bibliography{reference}

\end{document}